# Effects of dilute coal char particle suspensions on propagating methane detonation wave

Jingtai Shi[a,b], Pikai Zhang[b,c], Yong Xu[b], Wanxing Ren[a], Huangwei Zhang[b,*]

[a] *School of Safety Engineering, China University of Mining and Technology, Xuzhou, 221116, China*
[b] *Department of Mechanical Engineering, National University of Singapore, 9 Engineering Drive 1, Singapore 117576, Republic of Singapore*
[c] *National University of Singapore (Chongqing) Research Institute, Liangjiang New Area, Chongqing 401123, China*

**Abstract**

Methane/coal dust hybrid explosion is one of the common hazards in process and mining industries. In this study, methane detonation propagation in dilute coal char particle suspensions is studied based on Eulerian-Lagrangian method. The effects of char combustion on methane detonation dynamics are focused on. The results show that propagation of the methane detonation wave in coal particle suspensions are considerably affected by particle concentration and size. Detonation extinction occurs when the coal particle size is small and concentration is high. The averaged lead shock speed generally decreases with increased particle concentration and decreased particle size. Mean structure and interphase coupling of hybrid detonation are analysed, based on the gas and particle quantities. It is found that char combustion proceeds in the subsonic region behind the detonation wave and heat release is relatively distributed compared to that from gas phase reaction. The mass and energy transfer rates increase rapidly to the maximum near the reaction front in the induction zone. Moreover, for 1 μm particles, if the particle concentration is beyond a threshold value, detonation re-initiation occurs after it is quenched at the beginning of the coal dust suspensions. This is caused by hot spots from the shock focusing along the reaction front in a decoupled detonation and these shocks are generated from char combustion behind the lead shock. A regime map of detonation propagation and extinction is predicted. It is found that the re-initiation location decreases with the particle concentration and approaches a constant value when the concentration exceeds 1,000 g/m$^3$. Finally, the influence of coal particle surface reactions on gas chemistry under detonation relevant conditions is studied. It is found that the ignition delay time changes non-monotonically with particle size. The results from this study are useful for prevention and suppression of methane/coal dust hybrid detonation.

---

[*]Corresponding author. Tel.: +65 6516 2557; Fax: +65 6779 1459.
*E-mail address*: huangwei.zhang@nus.edu.sg.

# 1. Introduction

Methane/coal dust hybrid explosion is one of the common hazards in process and mining industries [1]. In the coal mine roadway, due to ventilation, fine coal particles may be suspended in the air. After they are heated by hot surrounding gas (e.g., from gas explosion), devolatilization and/or surface reaction can be initiated, through which volatile gas and reaction heat are released. This would considerably modulate the thermodynamic state of local flammable gas. Typically, existence of coal dust would complicate a gas explosion process and therefore make it more difficult to be predicted, compared to conventional gas explosion accidents [1]. Due to harsh experimental conditions and demanding requirement for modelling strategies to reproduce the multi-faceted physics, our understanding about combustion and explosion of methane and coal dust mixtures is still rather limited.

Investigations have been made about flammability limit, ignitability, and flame propagation in methane/coal dust two-phase mixtures. For instance, Cloney et al. [2] investigated the burning velocity and flame structures of hybrid mixtures of coal dust with methane below the lower flammability limit of the gaseous mixture. They correlated the unsteady flame behaviors (e.g., burning velocity oscillation) with combustion of volatile gas released from the dispersed particles. Xu et al. [3] found that both maximum explosion overpressure and overpressure rise rate increase with increased coal dust concentrations and decreased diameter. Xu et al. also studied the performance of mitigation of methane/coal dust explosion with fine water sprays [3–5]. Xie et al. [6] observed that flame burning velocity decreases when coal particle of sizes 53-63 µm and 75-90 µm are added, irrespective of the gas equivalence ratios. They also identified two competing effects associated with the volatile gas release (heat absorption, which corresponds to thermal effect) and addition (kinetic effect). Rockwell and Rangwala [7] found that turbulent burning velocity of methane flames increases as the coal particle size decreases and the concentration increases (>50%). This is in line with the findings by Chen [8], where he observed that presence of methane in coal dust explosions enhances the flame velocity of the mixture. Furthermore,

Amyotte et al. [9] studied the ignitability of methane/coal dust mixture and found that the apparent lean flammability limit decreases with high methane concentration, small particle diameter, and increased volatile matter content. Ma et al. [10] observe that the low-temperature oxidation of coal dust had prolonged the combustion process of methane-air/coal dust mixtures. Li et al. [11] experimentally investigated the influence of pre-oxidized status of coal dust on the deflagration severities and flame behaviors of methane–pulverized coal mixtures. They found that the pre-oxidization of coal dust would promote the explosion severity but prolong the burning time of the methane/coal dust mixtures.

In recent years, interactions between blast wave and coal dust are also studied based on multiphase numerical simulations. Houim et al. [12] studied the layered coal dust combustion induced by a blast wave degraded from a methane detonation. It is shown that the high-speed post-shock flow lifts the coal dust at the bottom of the domain, which ignites by a reaction wave of burning carbon char and generates a shock-flame complex. The coal-dust combustion generates pressure waves that overtake the lead shock and intensify the latter. In a subsequent study [13], they also found that inert layers of dust substantially reduce the overpressure, impulse, and speed produced by the propagating blast wave. The shock and flame are more strongly coupled for loose dust layers (initial volume fraction 1%), thereby propagating at a higher velocity and producing large overpressures. More recently, Guhathakurta and Houim observed that the role of heat radiation in layered dust explosions is affected by coal dust volume fraction [14]. With the similar configuration, Shimura et al. [15] numerically investigated the flame structure during shock - induced layered coal dust combustion an extended CFD–DEM model. They found that the dust particles mainly devolatilize behind the reaction front.

In the above work [12–15], since only incident blast wave is considered, how methane detonation interacts with the coal dust is not still clear. Moreover, for micro-sized coal dust, they may be easily aerosolized in the air by any aerodynamic perturbation. Therefore, it is necessary to understand how the coal dust suspensions affect an incident propagating detonation wave.

In this study, transmission of a methane detonation in dilute coal char particle suspensions will be simulated based on Eulerian-Lagrangian method. Two-dimensional configuration will be considered, and a reduced chemical mechanism [16] will be employed. The effects of coal particle concentration and size on methane detonation dynamics and parameters will be analyzed. The objectives of this work are to study: (1) the effects of coal particle suspensions on methane detonation speed; (2) detailed hybrid detonation structures; (3) mechanism of detonation extinction and re-initiation in coal particle suspensions; and (4) interphase coupling between the detonation wave and coal particles. The results from this work will be useful for understanding the fundamentals behind methane/coal dust hybrid detonation and providing the guidance for explosion inhibition measures in industrial practice. The manuscript is structured as below. The physical model will be introduced in Section 2, whilst the mathematical model is in Section 3. The simulation results and discussion will be given in Section 4, followed by the main conclusions in Section 5.

## 2. Physical model

Transmission of a methane detonation wave in dilute coal particle suspensions will be studied based on a two-dimensional configuration. Due to the reactive nature of the particles, this problem can be categorized into hybrid detonation, following Veyssiere [17]. The schematic of the physical model is shown in Fig. 1. The length ($x$-direction) and width ($y$) of the domain are 0.3 m and 0.025 m, respectively. It includes gaseous detonation development section (0−0.2 m) and gas-particulate two-phase section (0.2−0.3 m). The whole domain is initially filled with stoichiometric $CH_4/O_2/N_2$ (1:2:1.88 by vol.) mixture. The initial gas temperature and pressure are $T_0 = 300$ K and $p_0 = 50$ kPa, respectively.

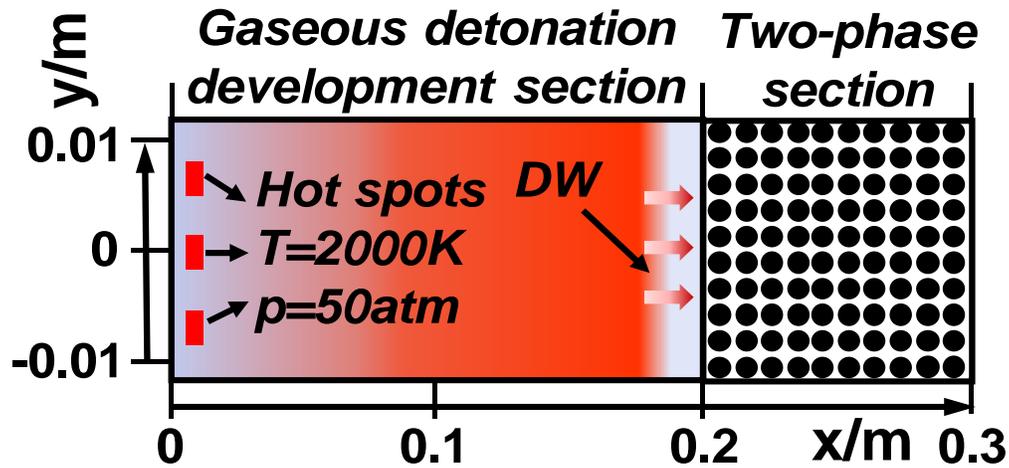

Fig. 1. Schematic of the physical problem. *x* and *y* axes not to scale.

In the two-phase section, coal char particles (for brevity we term as *coal particles* or simply *particles* hereafter) are uniformly distributed before the detonation arrives, to mimic coal dust suspensions in methane explosion hazards. In this study, the coal particle diameter varies from $d_p^0$ = 1 to 10 μm. The coal particle concentration ranges from $c$ = 10 to 1000 g/m$^3$. The resultant initial volume fractions are 0.0007%-0.067%, which are well below the upper limiting volume fraction (0.1%) of dilute particle-laden flows [18]. In our simulations, devolatilized coal particles are considered and therefore the devolatization process are excluded. This enables us to concentrate on the effects of dispersed coal char combustion on methane detonation dynamics, which is the objective of this study. The particle is composed of inert ash and fixed carbon, with mass fractions of 11.3% and 88.7%, respectively. The heat capacity and initial material density of the particle are 710 J kg$^{-1}$ K$^{-1}$ and 1,500 kg m$^{-3}$. These properties approximately follow the properties of typical bituminous coals [19].

The detonation wave (DW) is initiated by three hot spots (2 mm × 4 mm, 2,000 K and 50 atm, see Fig. 1) near the left end of the domain. The gaseous detonation development section (0.2 m) is sufficiently long to achieve a freely propagating methane DW before it enters the two-phase section. For all gas-particulate two-phase detonations simulations, a consistent initial field with propagating detonation wave at about $x = 0.196$ m (i.e., slightly before the two-phase section) is used. Therefore, two-phase simulations only run from $x = 0.196$ m to 0.3 m.

The upper and lower boundaries of the domain in Fig. 1 are periodic. At the left boundary of the domain, $x = 0$, non-reflective condition is enforced for the pressure, whereas zero-gradient condition for other quantities. At $x = 0.3$ m, zero-gradient conditions are employed for all variables.

Uniform Cartesian cells are used to discretize the domain in Fig. 1 and the CFD mesh size is 50 μm at $x = 0$–0.14 m and 25 μm at 0.14–0.3 m. The resultant cell numbers are 7,800,000. The same mesh resolution is used in our recent work on methane detonation inhibition by fine water mists [20]. The Half-Reaction Length (HRL) from the ZND structure of the stoichiometric $CH_4/O_2/N_2$ mixture is about 2,200 μm under the simulated conditions. Thus, for the two-phase section where our study is performed, there are about 88 cells within the HRL of C-J detonation. We also perform the mesh sensitivity analysis through halving the resolution in the two-phase section (12.5 μm). The results are provided in section A of the supplementary document, which show that the cell regularity and size of the detonation predicted with the two resolutions are generally close.

## 3. Mathematical model

### 3.1 Governing equation

The Eulerian-Lagrangian method is used for simulating methane/coal char particle hybrid detonations. For the continuous phase, the conservation laws of mass, momentum, energy, and species mass fraction are solved for the multi-species, compressible, reacting flows. The equations read

$$\frac{\partial \rho}{\partial t} + \nabla \cdot [\rho \mathbf{u}] = S_{mass}, \qquad (1)$$

$$\frac{\partial (\rho \mathbf{u})}{\partial t} + \nabla \cdot [\mathbf{u}(\rho \mathbf{u})] + \nabla p + \nabla \cdot \mathbf{T} = \mathbf{S}_{mom}, \qquad (2)$$

$$\frac{\partial(\rho E)}{\partial t} + \nabla \cdot [\mathbf{u}(\rho E + p)] + \nabla \cdot [\mathbf{T} \cdot \mathbf{u}] + \nabla \cdot \mathbf{j} = \dot{\omega}_T + Q_{g,rad} + S_{energy}, \quad (3)$$

$$\frac{\partial(\rho Y_m)}{\partial t} + \nabla \cdot [\mathbf{u}(\rho Y_m)] + \nabla \cdot \mathbf{s_m} = \dot{\omega}_m + S_{species,m}, (m = 1, \ldots M - 1). \quad (4)$$

In the above equations, $t$ is time and $\nabla \cdot (\cdot)$ is the divergence operator. $\rho$ is the gas density, and $\mathbf{u}$ is the gas velocity vector. $p$ is the pressure, and is from the ideal gas equation of state, i.e., $p = \rho R T$. $T$ is the gas temperature. $R$ is the specific gas constant. And is calculated from $R = R_u \sum_{m=1}^{M} Y_m W_m^{-1}$. $W_m$ is the molar weight of $m$-th species and $R_u = 8.314$ J/(mol·K) is the universal gas constant. In Eq. (2), $\mathbf{T}$ is the viscous stress tensor. In Eq. (3), $\mathbf{j}$ is the diffusive heat flux and $E$ is the total non-chemical energy. Also, the term $\dot{\omega}_T$ represents the heat release rate from the chemical reactions. In Eq. (4), $Y_m$ is the mass fraction of $m$-th species, and $M$ is the total species number. $\mathbf{s_m}$ is the species mass flux, and $\dot{\omega}_m$ is the reaction rate of $m$-th species by all $N$ reactions.

In this study, a reduced methane mechanism (DRM 22) [16] is used, including 24 species and 104 reactions. The accuracy of the DRM 22 mechanism in detonation simulations has been evaluated by Wang et al. [21], including the ignition delay time over a range of operating conditions. In section B of the supplementary document, we also compare the C-J speed and pressure/temperature at the C-J and von Neumann points of the ZND structure predicted with DRM 22 and GRI 3.0 [22] and find that the results from them are similar.

The Discrete Ordinate Method (DOM) is based on discretizing the direction of radiation intensity ($I_i$) and solving the radiation transfer equation in the discrete direction of solid angle [23]. In the 2-D rectangular coordinate system, the radiative transfer equation can be written as [23]

$$\mu_i \frac{\partial I_i}{\partial x} + \eta_i \frac{\partial I_i}{\partial y} = \kappa(I_b - I_i), \quad (5)$$

where $I_i [\equiv I(x, y; \Omega_i)]$ is radiation intensity at position $(x, y)$ in the discrete direction $\Omega_i$, $\mu_i$ and $\eta_i$ are directional cosine, $I_b$ is the blackbody radiation intensity. The Planck mean absorption coefficient of the gas mixture is

$$\kappa = \sum_i k_i p_i, \tag{6}$$

where $\kappa_i$ and $p_i$ are the Planck mean absorption coefficient and partial pressure of $i$-th species, respectively. In the present study, $CO_2$, $CO$, $CH_4$, and $H_2O$ are the radiant species and the mean absorption coefficients are taken from Refs. [24,25].

The gas-phase radiation heat transfer term in the energy equation, Eq. (3), reads

$$Q_{g,rad} = -\kappa \left( 4\pi I_b - \int_{4\pi} I \, d\Omega \right), \tag{7}$$

where $\int_{4\pi} I \, d\Omega$ is the incident radiation intensity and $\Omega$ is the solid angle. It can be further written as the following form when $I_i$ is solved from DOM

$$Q_{g,rad} = -\kappa \left( 4\sigma T^4 - \sum_i \omega_i I_i \right), \tag{8}$$

where $\sigma$ is the Stephen–Boltzmann constant, $\omega_i$ is the weight for the $i$-th ordinate, and $\sum_i \omega_i I_i = \int_{4\pi} I \, d\Omega$.

For the coal particle phase, the Lagrangian method is used to track coal particles. Particle collisions are neglected because the collision timescale is much longer than the momentum relaxation timescale when the particle concentration is dilute [18]. It is assumed that the temperature is uniform inside the particles due to their low Biot numbers of coal particles (<0.005). Gravitational force is not included due to smallness of the particles. Coal particles are assumed to spherical, and the swelling effect is not considered. Therefore, the particle diameter is constant throughout the simulations. With above assumptions, the evolutions of mass, momentum, and energy of a coal particle are governed by

$$\frac{dm_p}{dt} = -\dot{m}_p, \tag{9}$$

$$m_p \frac{d\mathbf{u}_p}{dt} = \mathbf{F}_d + \mathbf{F}_p, \tag{10}$$

$$m_p c_{p,p} \frac{dT_p}{dt} = \dot{Q}_s + \dot{Q}_c - Q_{p,rad} + Q_{g,rad-p}, \tag{11}$$

where $m_p = \pi \rho_p d_p^3/6$ is the mass of a single particle, $\rho_p$ and $d_p$ are the particle material density and diameter, respectively. $\dot{m}_p$ is the surface reaction rate and $\mathbf{u}_p$ is the particle velocity vector. The Stokes drag force is modelled as [26]

$$\mathbf{F}_d = (18\mu/\rho_p d_p^2)(C_d Re_p/24) m_p (\mathbf{u} - \mathbf{u}_p), \tag{12}$$

while the pressure gradient force or Archimedes force is

$$\mathbf{F}_p = -V_p \nabla p. \tag{13}$$

Here $V_p$ the volume of a particle.

In Eq.(12), the drag coefficient, $C_d$, is estimated with [26]

$$C_d = \begin{cases} 0.424, & \text{if } Re_p \geq 1000, \\ \dfrac{24}{Re_p}\left(1 + \dfrac{1}{6} Re_p^{2/3}\right), & \text{if } Re_p < 1000. \end{cases} \tag{14}$$

The particle Reynolds number, $Re_p$, is defined as

$$Re_p \equiv \frac{\rho d_p |\mathbf{u}_p - \mathbf{u}|}{\mu}. \tag{15}$$

The char combustion (or particle surface reaction) is modelled using a global reaction, $C_{(s)} + O_2 \rightarrow CO_2$, where $C_{(s)}$ is fixed carbon. The kinetic/diffusion-limited rate model [27] is used to estimate the reaction rate, i.e.,

$$\dot{m}_p = A_p p_{ox} D_0 R_k / (D_0 + R_k), \tag{16}$$

which accounts for the particle mass change in Eq. (9). $p_{ox}$ is the partial pressure of oxidant species in the surrounding gas. The diffusion rate coefficient $D_0$ and kinetic rate coefficient $R_k$ are respectively estimated from

$$D_0 = C_1 \left[(T + T_p)/2\right]^{0.75} / d_p, \tag{17}$$

$$R_k = C_2, \tag{18}$$

The model constants $C_1$ and $C_2$ are $5\times10^{-12}$ kg/(m·s·Pa·K$^{0.75}$) and 0.002 kg/(m$^2$·s·Pa), respectively, whilst the activation energy $E$ is $7.9\times10^7$ J/kmol [28,29].

In Eq. (11), $c_{p,p}$ is the particle heat capacity and $T_p$ is the particle temperature. $\dot{Q}_s$ is the rate of char combustion heat release absorbed by the particle. The convective heat transfer rate is

$$\dot{Q}_c = h_c A_p (T - T_p), \tag{19}$$

where $A_p$ is the particle surface area. $h_c$ is the convective heat transfer coefficient, estimated with the Ranz and Marshall correlation [30].

Moreover, the radiative emission rate from a particle reads

$$Q_{p,rad} = A_p \varepsilon_p \sigma T_p^4, \tag{20}$$

where $\varepsilon_p$ is the emissivity of particle surface and is assumed to unity because the major composition is carbon [31]. The particle radiation absorption rate takes the following form

$$Q_{g,rad-p} = A_p \varepsilon_p \sum_i \omega_i I_i / 4. \tag{21}$$

The effects of coal particles on the gas phase are considered through the Particle-source-in-cell approach [32], in terms of the mass, momentum, energy and species exchanges. These respectively correspond to the source terms in the Eqs. (1)-(4), i.e. $S_{mass}$, $\mathbf{S}_{mom}$, $S_{energy}$ and $S_{species,m}$, can be estimated based on the particles in individual CFD cells, which read ($V_c$ is the cell volume, $N_p$ is the particle number in the cell)

$$S_{mass} = \frac{1}{V_c} \sum_1^{N_p} \dot{m}_p, \tag{22}$$

$$\mathbf{S}_{mom} = -\frac{1}{V_c} \sum_1^{N_p} (\mathbf{F}_d + \mathbf{F}_p), \tag{23}$$

$$S_{energy} = -\frac{1}{V_c} \sum_1^{N_p} (\dot{Q}_s + \dot{Q}_c), \tag{24}$$

$$S_{species,m} = \begin{cases} -\dfrac{W_{O2}}{W_C} S_{mass} & for\ O_2\ species, \\ \dfrac{W_{CO2}}{W_C} S_{mass} & for\ CO_2\ species, \\ 0 & for\ other\ species. \end{cases} \qquad (25)$$

**3.2 Numerical method and solver validation**

The gas and particulate phase equations are solved using an OpenFOAM code for two-phase compressible reacting flow, *RYrhoCentralFOAM* [33–37]. Details about the numerical methods in *RYrhoCentralFOAM* can be found in Refs. [34,38], and in this section only key information is presented for completeness.

For the gas phase equations, second-order backward scheme is employed for temporal discretization and the time step is about $9\times10^{-10}$ s. A MUSCL-type and Riemann-solver-free scheme [39] with van Leer limiter is used for convective flux discretization in the momentum equations. Total variation diminishing scheme is used for the convection terms in the energy and species equations. Second-order central differencing is applied for all diffusion terms.

For the particulate phase, Eqs. (9)-(11) are integrated with an Euler implicit method and the right-side terms are treated in a semi-implicit fashion. Computational parcel concept is used and one parcel denotes ensemble of coal particles with identical properties, such as diameter, size, temperature and velocity [18,40,41]. In our simulations, the number of parcels distributed in the two-phase section is about 5 million, and the coal particle number in a parcel is varied based on the particle size and concentration.

The solver *RYrhoCentralFOAM* has been extensively validated and verified in our previous studies[34–36], in terms of shock-capturing, molecular diffusion, flame-chemistry interactions, and two-phase gas-droplet coupling. Here we further validate the solver against the shock-particle interaction experiments by Sommerfeld [42]. In this experiment, a shock wave of Mach 1.49 propagates into a particle-laden area. The particles are spherical glass beads, and the material density, heat capacity and mean diameter are 2.5 g/cm$^3$, 766 J/kg/K, and 27 μm, respectively. Two particle volume fractions are considered, i.e., $\alpha_p$ = 0.0249% and 0.0584%, which are close to the upper limit of the particle volume fractions studied in this work. Figure 2 shows that our solver can accurately reproduce the evolutions of the shock Mach number subject to the dispersed particles with different volume fractions, which further corroborates the solver accuracy for predicting gas-particulate two-phase flows.

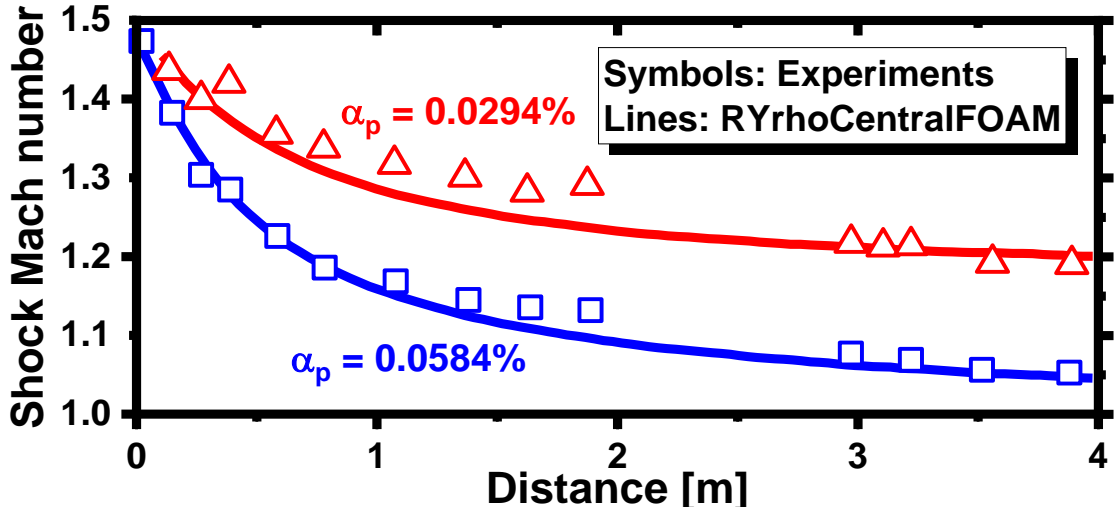

Fig. 2. Evolutions of shock Mach number in shock-particle interactions. Experimental data: Ref. [42].

## 4. Results and discussion

### 4.1 Detonation propagation speed

Plotted in Fig. 3 is the average lead shock speed, $\bar{D}$, as a function of particle concentrations for different particle diameters. $\bar{D}$ is calculated from the length of two-phase section (i.e., 0.1 m) divided by the total shock residence time in this section. When the particle concentrations are low, e.g., 10 and 50 g/m$^3$ as shown in the inset of Fig. 3, $\bar{D}$ is slightly higher (by 2%) than that of the purely gaseous case ($c$ = 0, pink square) for all four diameters. This means that in coal dust suspension with small particle concentrations, shock transmission speed is enhanced, compared to that in gas-only mixture. This is because of quickly excited surface reactions of the particle in the detonated gas [12,13]. Fast energy deposition into the gas phase can emanate the right-running pressure or shock waves, thereby intensifying the lead shock. This is termed as *surface reaction effects* hereafter. This speed enhancement by reactive dispersed particles (e.g., wheat particles) is also observed in Refs. [17,43,44] in their studies.

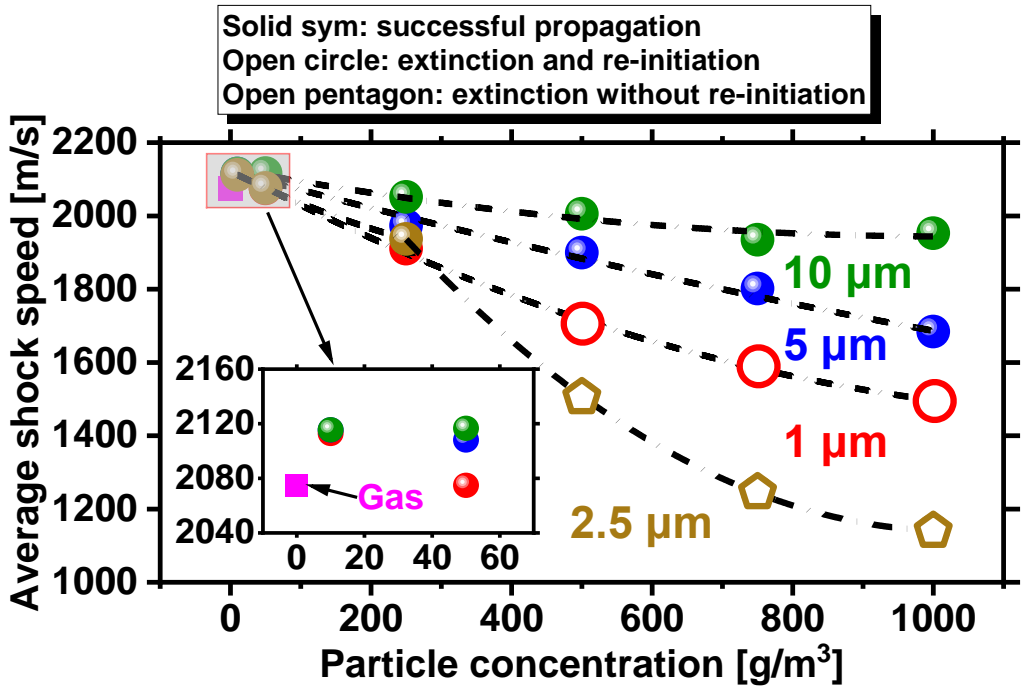

Fig. 3. Change of averaged lead shock speed as a function of coal particle concentration for different particle sizes.

We further look at the cases with higher concentrations, beyond 50 g/m$^3$, in which cases the particle diameter influences are pronounced. Specifically, for 5 and 10 µm particles, detonation transmission is observed and $\bar{D}$ decreases when $c$ is increased. The latter is because for higher concentration of particles more energy / momentum would be absorbed from the gas to heat / accelerate themselves. Therefore, the shock intensity is reduced as $c$ increases. We term this as *energy / momentum effects* for brevity, which dominate over the surface reaction effects in these cases.

Nonetheless, for 2.5 µm coal particles and $c$ = 500-1000 g/m$^3$, detonation extinction occurs once they arrive at the coal suspension area. Therefore, the average shock speed of these cases (open pentagons in Fig. 3) is much lower than the others. In these cases, the energy / momentum effects become more significant. Nonetheless, for 1 µm particles with high concentration (i.e., > 500 g/m$^3$), the DW is quenched when they enter the two-phase section, same as the 2.5 µm cases. However, different from 2.5 µm particles, detonation is re-initiated due to the particle surface reactions, and the shock speed averaged from the shock residence time in the two-phase section is therefore generally higher than those of 2.5 µm. The transient and mechanism of DW re-initiation by surface reaction will be further interpreted in Section 4.4.

**4.2 Particle concentration effects**

Figure 4 shows the peak pressure trajectories for methane detonations with various coal particle concentrations (10-1000 g/m$^3$). The particle diameter is fixed to be $d_p$ = 1 µm. The results from particle-free case ($c$ = 0) are illustrated in Fig. 4(a) for comparison. One can see that coal particle suspensions considerably change the cellular structures of methane detonations. Specifically, when $c$ = 10 g/m$^3$, the cells are generally regular and small, which are particularly pronounced for the second half of the two-phase section ($x$ > 0.25 m), through comparisons with the particle-free case. This indicates that more stable propagation occurs when relatively dilute particles are loaded, consistent with the enhanced speed revealed in Fig. 3. This is because the combustible coal particles provide additional heat release from the surface reactions, generating pressure impulse towards the lead shock and hence enhancing the frontal stability [12,13,17]. When $c$ is further increased, e.g., 50 and 250 g/m$^3$ in Figs. 4(c) and 4(d), the cell size (apex-to-apex distance, $\lambda$, as annotated in Fig. 4d) generally increases, with the mean cell widths being about 5.6 and 12.5 mm, respectively.

However, when $c$ = 500 and 1000 g/m$^3$, the DW extinction occurs when it just encroaches the coal particle area. This is caused by the strong effects of energy and mentum absorption effects by the coal particle; meanwhile, the surface reactions have not started due to particle heating period and finite-rate surface reaction kinetics. From Figs. 4(e) and 4(f), the overpressures are significantly reduced, which is because the reactivity of the triple points (where the trajectories are mostly from) is highly reduced due to the decoupling of reactive front and lead shock. Nonetheless, the detonation is re-initiated downstream in the particle suspensions, e.g., at $x = 0.23$ m when $c = 1000$ g/m³. This is accompanied by sudden intensification of local pressures, as marked as several discrete high-pressure spots (HPS) in both Figs. 4(e) and 4(f). The sudden pressure rise at these locations is caused by localized explosion with the nature of isochoric combustion. The whole re-initiation process can also be watched from the videos submitted with this manuscript. Onset of these HPS is a significant feature in methane and fine coal particle hybrid detonations. In practical explosion hazards, it may induce secondary detonation or explosion, thereby increasing g the severity of the consequences.

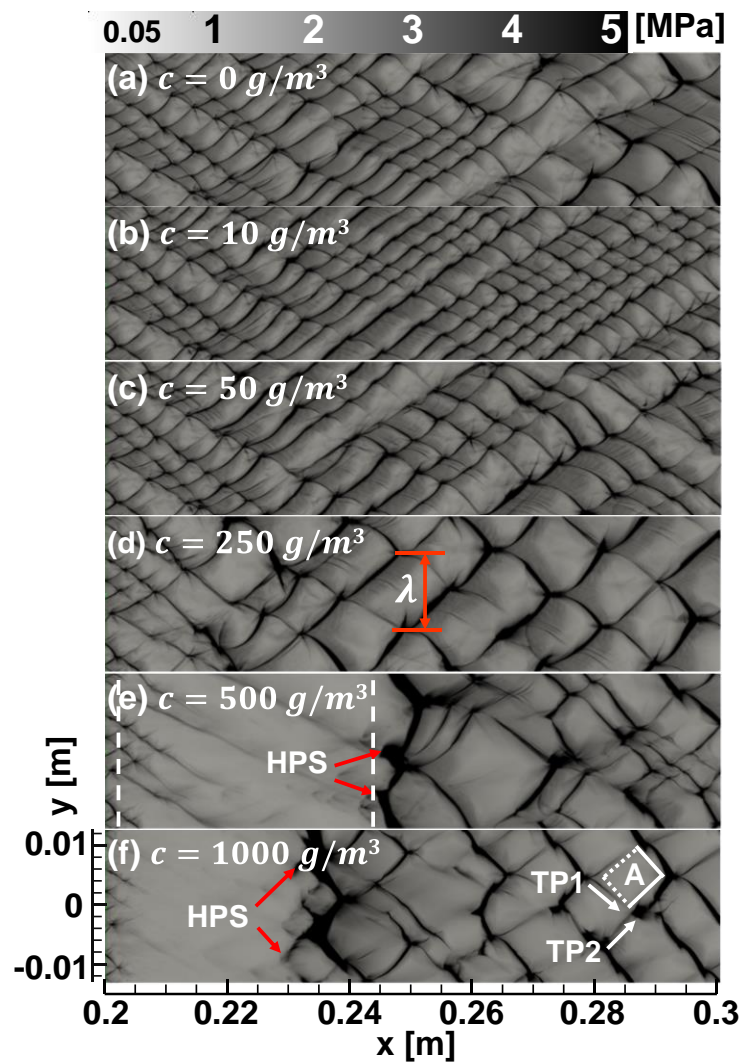

Fig. 4. Peak pressure trajectories with different coal particle concentrations. $d_p = 1$ μm. TP: triple point; HPS: high-pressure spot.

Further downstream, clear cells appear again, but the strength of the triple point trajectories in one cell of the re-initiated detonations is different. This leads to a different morphology (wave-like along the width) of peak pressure trajectories beyond $x = 0.25$ m from Figs. 4(c) and 4(d), and this is more notable in Fig. 4(f). For instance, in cell A marked in Fig. 4(f), the weak and strong trajectories are highlighted with dashed and solid lines, respectively. The weaker trajectories are caused by the decoupling of the reaction fronts from the weakened Mach stem due to the foregoing energy-momentum effects (see section D of the supplementary document and the videos). After two triple points (TP1 and TP2) collide, a new Mach stem is formed, and the pressure peak trajectory is strengthened (solid edges of cell A).

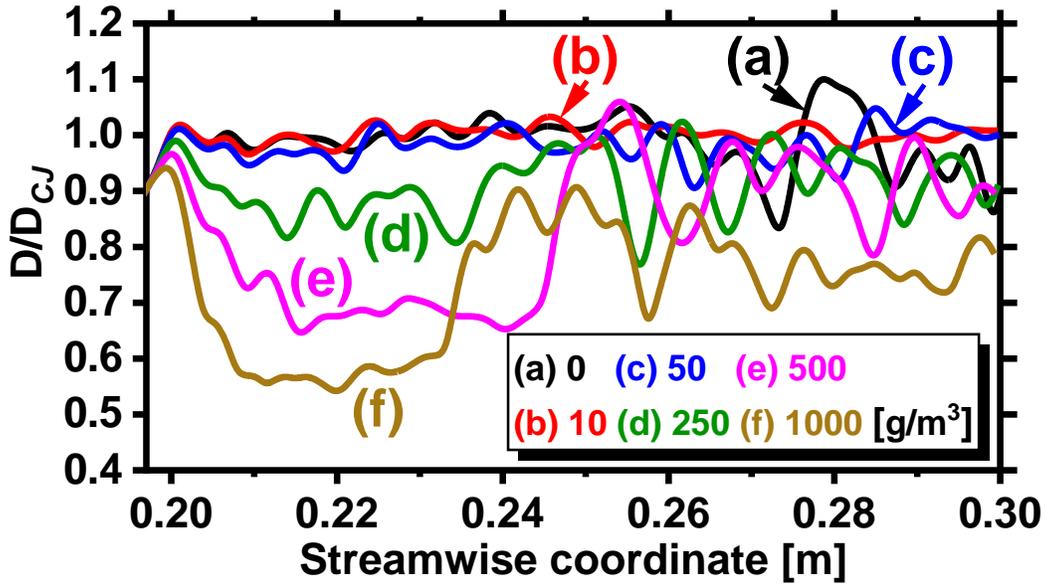

Fig. 5. Evolutions of lead shock speed with various coal char particle concentrations. $D_{CJ}$ is the Chapman–Jouguet speed (2,109 m/s) for particle-free $CH_4/O_2/N_2$ mixture.

Figure 5 shows the evolutions of lead shock propagation speed $D$ in the two-phase section (0.2-0.3 m) for six cases in Fig. 4. Note that they are calculated from the timeseries of lead shock positions with a time interval of one microsecond. As demonstrated from lines b and c, when $c = 10$ and 50 g/m$^3$, the shock speeds $D$ fluctuate around $D_{CJ}$, which is the C-J speed of the particle-free $CH_4/O_2/N_2$ mixture. They are also close to the results of the gas-only case, i.e., line a in Fig. 5(a). However, with $c = 250$ g/m$^3$, the DW has generally lower speed with stronger fluctuations. This is caused by stronger energy / momentum effects by higher loading of coal char particles. For lines e and f, due to higher particle concentrations, the lead shock speed is considerably reduced to around 70% and 55% of $D_{CJ}$, respectively, in the first half of the two-phase section. This can be justified by the decoupling of reactive front from the lead shock wave, as evidenced in Figs. 4(e) and 4(f). Nonetheless, for $x \geq 0.24$ m, since detonation re-initiation occurs, the lead shock speeds are quickly restored, but still well below the C-J speed (generally 70%-80% $D_{CJ}$) due to the strong interphase exchanges between the gas and fine particles.

**4.3 Particle diameter effects**

Figure 6 shows the DW peak pressure trajectories with various coal particle diameters (1-10 μm). The particle concentration is $c = 500$ g/m³. One can see that coal particle sizes exhibit significant effects on detonation propagation in the coal particle suspension. When the particle size is large, e.g., $d_p = 10$ and 5 μm in Figs. 6(a) and 6(b), DW transmission can occur. Nonetheless, the average cell size generally increases in the second half of the two-phase section ($x > 0.25$ m) and the cells become more irregular, compared to the gas-only results in Fig. 4(a). This indicates more unstable DWs due to the effects of the dispersed coal particles. In Fig. 6(c), for $d_p = 2.5$ μm, the DWs can propagate a distance in the particle suspensions, but the peak pressures are considerably reduced beyond $x = 0.22$ m. This indicates the occurrence of the detonation extinction and the exchanges of mass, momentum, and energy between them will be discussed in Section 4.5. Furthering decreasing the particle size to 1 μm results in detonation extinction and re-initiation, as already discussed in Fig. 4(e).

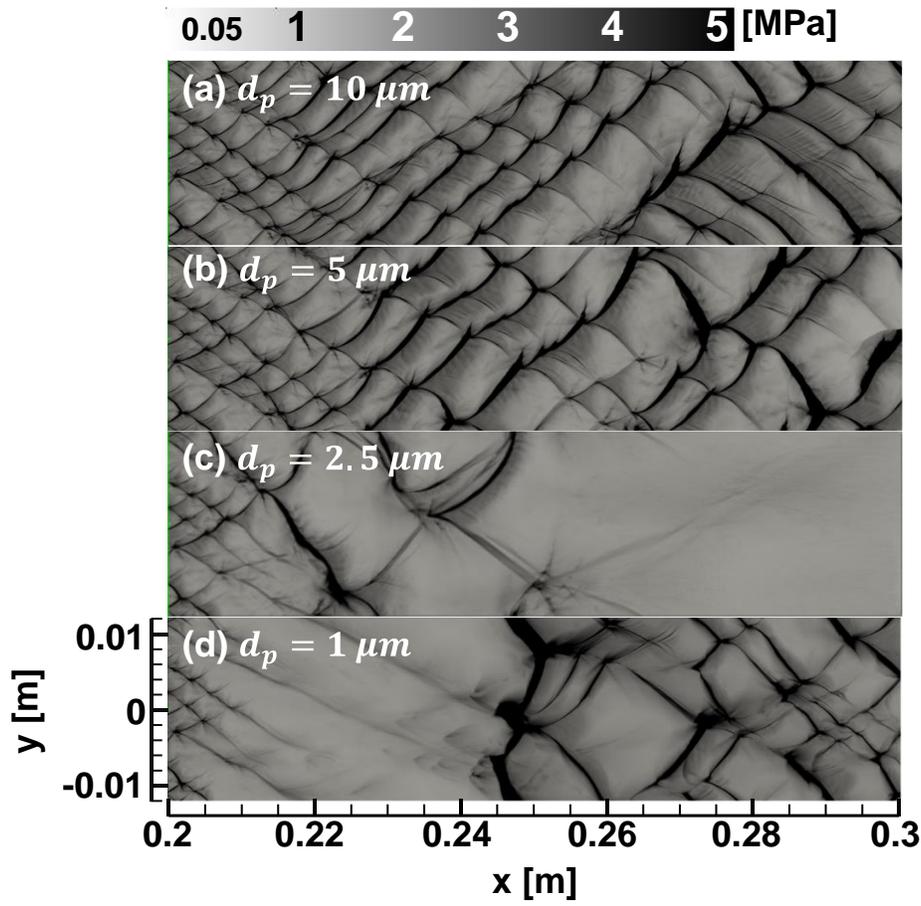

Fig. 6. Peak pressure trajectories with different coal particle sizes. $c = 500$ g/m$^3$.

Figure 7 further shows the spatial evolution of lead shock speed in four cases in Fig. 6. With increased $d_p$ from 1 μm to 10 μm, the lead shock speeds are generally reduced after the DW enters the suspension at $x = 0.2$ m. The curve of $d_p = 1$ μm has been interpreted in Fig. 5, but added here for comparison. When $d_p = 10$ μm, the speed fluctuates little around the C-J speed, which implies that the particles of the size have relatively small effects on the detonation transmission. For $d_p = 5$ μm, the speed fluctuation is more obvious beyond 0.24 m. This shows that the surface reaction of coal particles is more intense in the second half of the two-phase section. For $d_p = 2.5$ μm, since DW decoupling occurs, the speed decreases, well below the C-J value. Moreover, in the first half of the two-phase section (0.2-0.25 m), it can be seen that the larger the particle size is, the smaller the speed attenuation is. This is because the larger the particle size, the less heat released by the surface reaction of the particles, and thus the weaker the effect on the detonation wave. This can well justify why the detonation speed is relatively higher when particle is coarser.

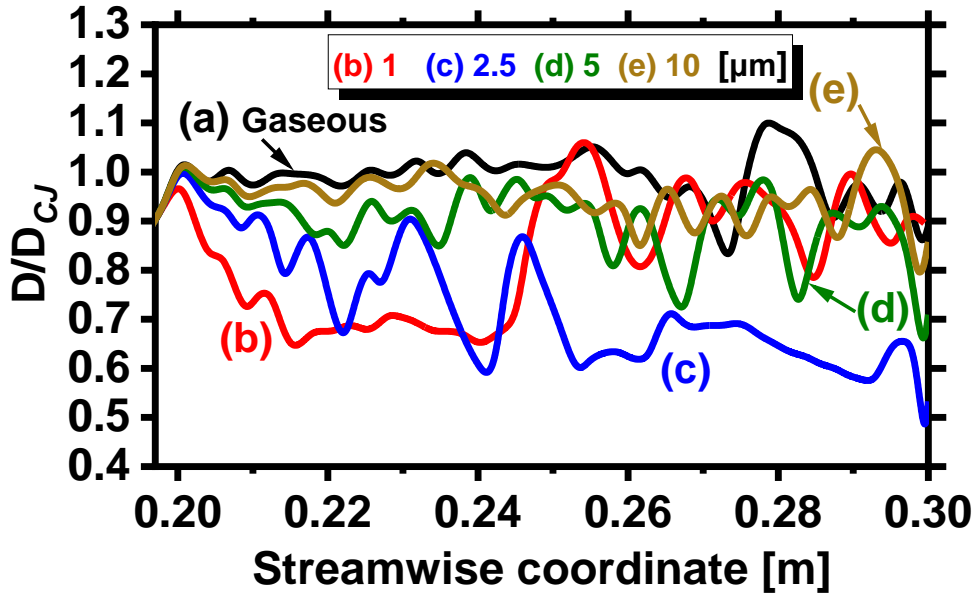

Fig. 7. Evolutions of lead shock speed with various coal char particle sizes. $D_{CJ}$ is the Chapman–Jouguet speed (2,109 m/s) for particle-free $CH_4/O_2/N_2$ mixture.

**4.4 Phenomenological description of detonation extinction and re-initiation**

It has been shown from Figs. 3, 4 and 6 that when the coal particle diameter is small (e.g., 1 µm) and concentration is high, unsteady detonation phenomena, e.g., extinction or re-initiation, would occur. To further elaborate, the results correspond to Fig. 4(f), i.e., $c$ = 1000 g/m³ and $d_p$ = 1 µm, will be discussed here. Figure 8 shows the time evolutions of gas temperature at four instants. Note that in our subsequent analysis, $t$ = 0 corresponds to the instant when $x$ = 0.196 m, i.e., immediately ahead of the two-phase section. At 3 µs, when the DW just enters the coal suspension, it is weakly unstable with multiple heads. However, at 5 µs, the distance between the lead shock front (LSF) and reaction front (RF) is increased, and the post-shock temperature is reduced to well below 2,000 K. Afterwards, the LSF and RF are completely separated, and therefore the detonative combustion extinguishes. This indicates that considerable energy is extracted from the gas to heat the coal particles and hence coupling between the shock and reaction front for detonative combustion cannot maintain.

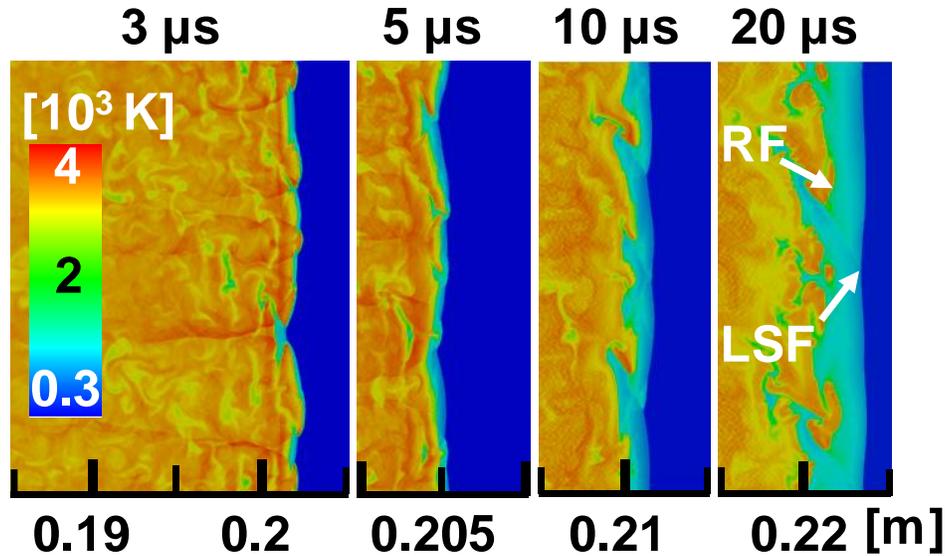

Fig.8. Time sequence of gas temperature in a detonation extinction. Tick spacing: 5 mm. $c = 1000$ g/m$^3$, $d_p = 1$ μm.

After the shock wave propagates a distance in the coal particle suspensions, re-initiation occurs, as shown in Fig. 4(f). Some instants of this transient are shown in Fig. 9, which are the continued development of the events in Fig. 8. Some evolving hot spots of different sizes appear along the RF, which are numbered as 1-3 in Fig. 9. They are characterized by locally elevated pressure (see the red colour in Fig. 10a), indicating the nature of isochoric combustion caused by the coherent interplay between strong heat release and pressure waves. These heat release points are gradually amplified as the carbon particles burn to form new shock waves. The HPS as shown in Fig. 4(f). The RF from them gradually catches up with the LSF, and therefore detonation along the LSF is intensified. Moreover, the blast waves from various chemical heat release locations collide, there by generating a high-temperature and high-pressure area (e.g., point 1 and 2 at 31 μs), which accelerates the occurrence of re-initiation.

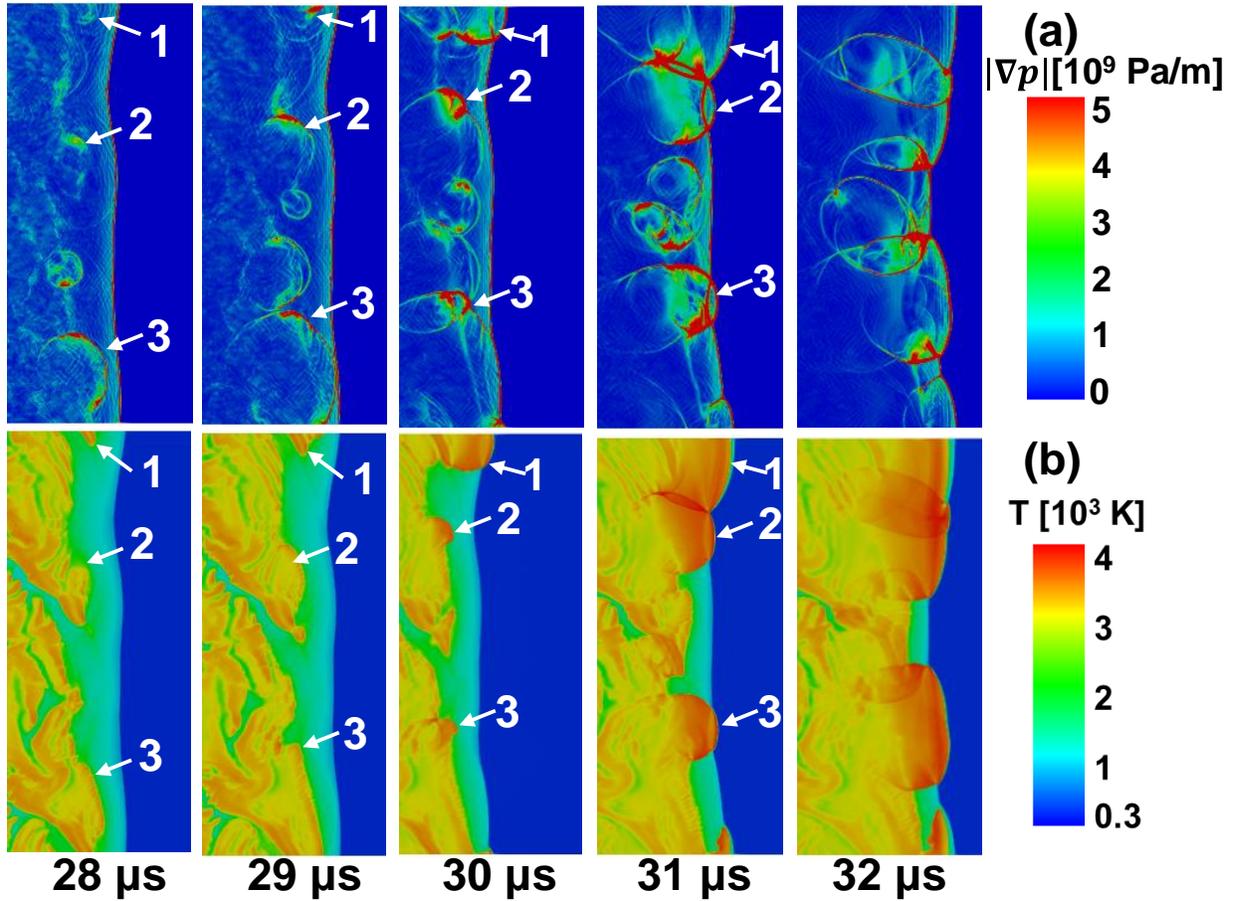

Fig. 9. Distributions of (a) pressure gradient magnitude and (b) gas temperature when the detonation is re-initiated. 1, 2 and 3: hot spots. $c = 1000$ g/m$^3$.

The flow structure behind the LSF can be clearly found from the pressure gradient magnitude in Fig. 9(a). The arched shocks are observable, which originate from spatially nonuniform surface reaction heat release from dispersed particles in the post-shock subsonic zones. We also perform a test with surface reaction de-activated for this case and find that there are no curved shocks behind the lead shock, and no re-initiation occurs either (see section C of the supplementary document). The propagation of these shocks results in the following unsteady events: (1) the forward-running components overtake and hence intensify the LSF; (2) the spanwise components re-compress the shocked gas and coal particles behind the LSF; and (3) more importantly, shock-focusing along the RF by these shocks leads to the formation of small reactive spots (e.g., 2 and 3). These spots quickly grow logitudinally and spanwise in the form of propagating reaction fronts, as evidenced in the results of 31 and 32 µs. Their leading sections overtake the LSF, which generates an overdriven Mach stem with strong gas reaction HRR (see Fig. 10b). The spanwise component evolves into the transverse wave extending from the triple pionts of the new MS (see 32 µs results). As such, the number of the new DW heads is largely correlated to the number of the hot spot and therefore randomness exists. This randomness comes from the inducing factors for hot spot formation, e.g., heterogeneous reaction, shock focusing location, and chemistry-shock interaction.

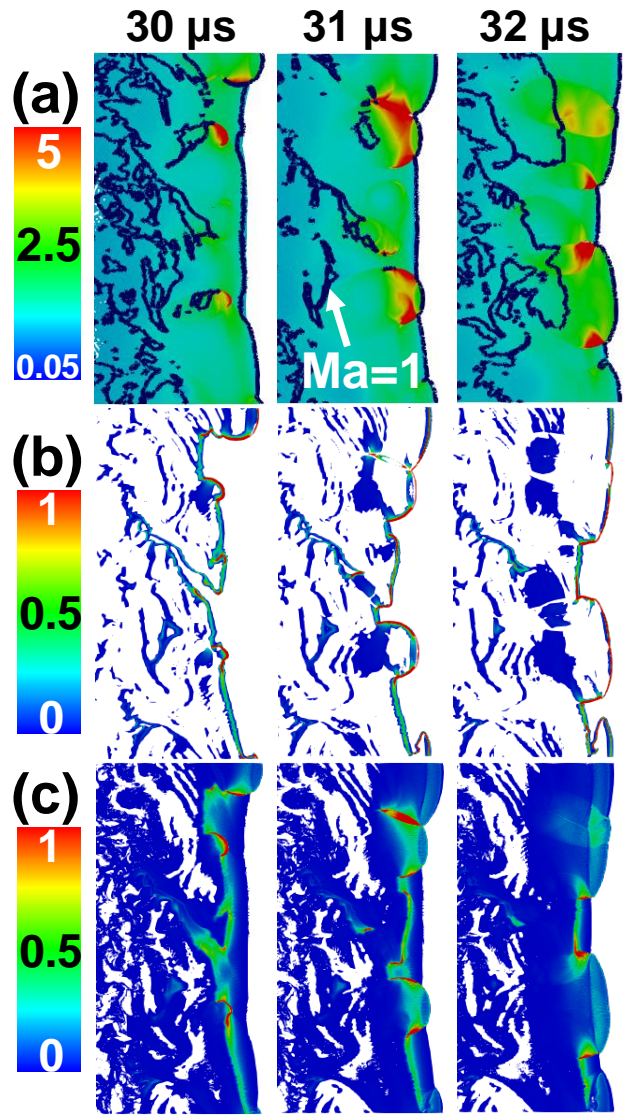

Fig. 10. Distributions of (a) pressure (in MPa), (b) gas reaction HRR ($10^{13}$ J/m$^3$/s), (c) surface reaction HRR ($10^{13}$ J/m$^3$/s) in a re-initiation process. 1, 2 and 3: hot spots. $c = 1000$ g/m$^3$ and $d_p = 1$ µm.

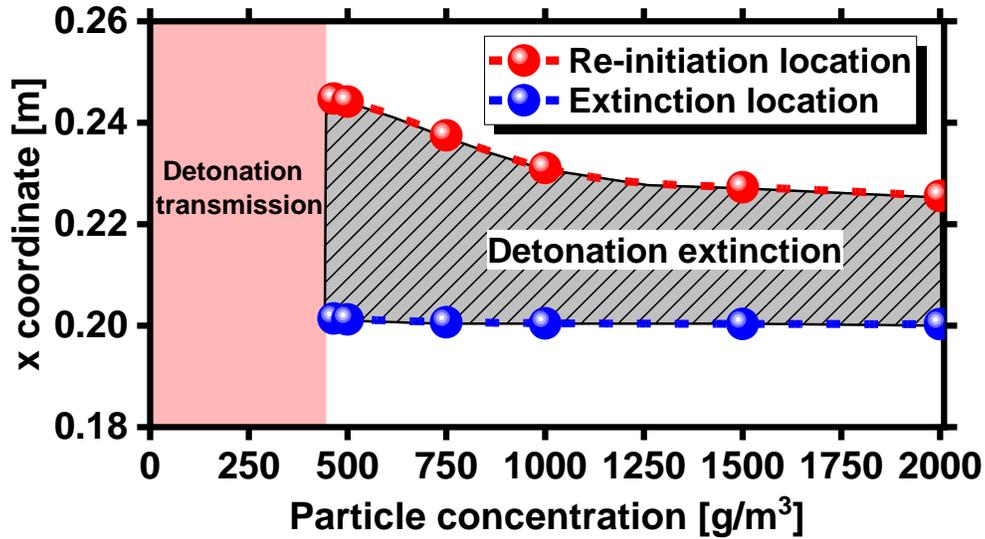

Fig. 11. Regime map of extinction and re-initiation with different particle concentrations. $d_p = 1$ µm.

The streamwise locations of detonation extinction and re-initiation under different coal particle concentration conditions are shown in Fig. 11. To reiterate, in our simulated cases, re-initiation phenomenon is only observed when $d_p = 1$ µm. The critical extinction (or re-initiation) location is determined from the *x*-coordinate where the peak pressure is critically lower than (or exceeds) 2.5 MPa, as shown with the two dashed lines in Fig. 4(e). This threshold is the average value of detonation wave maintaining stable propagation in these cases (see section E of the supplementary document). Slightly changing it would not induce qualitative differences in Fig. 11. When $c \leq 465$ g/m$^3$, the incident DW wave can successfully propagate in the coal particle suspensions (see the pink area in Fig. 11). Detonation extinction and re-initiation only occur when $c > 465$ g/m$^3$. Moreover, the critical extinction locations (blue line) are not sensitive to the coal particle concentration. It is around 0.2 m, indicating that extinction occurs almost immediately when the DW arrives at the particle suspensions. This is reasonable since fine coal particles have larger specific surface areas to have the energy and mass transfer between the continuous phase and particulate phase. Moreover, the re-initiation location decreases with the particle concentration. This is justifiable because higher concentration of coal particles leads to greater interphase exchanges of momentum and energy. However, as the particle concentration exceeds 1,000 g/m$^3$, the re-initiation location approaches a constant value of 0.225 m. This may be limited by the timescales of coal particle heating and/or surface reaction kinetics. Understanding the re-initiation distance is significant to prevent secondary explosion in real-world situations, e.g., for adding explosion suppressants (such as ultra-fine water mist) at the possible re-detonation locations.

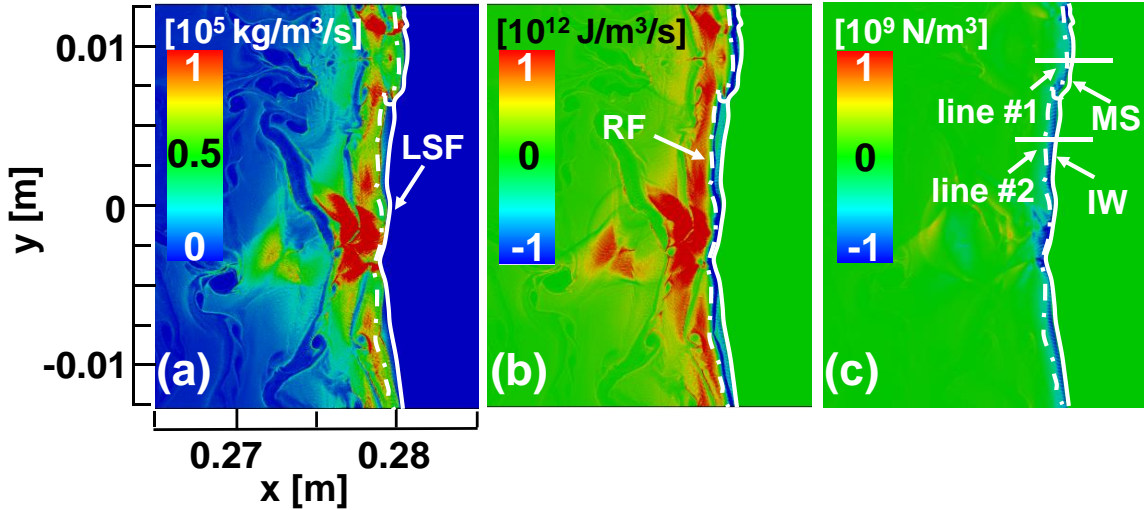

Fig. 12. Contours of (a) mass, (b) energy, and (c) momentum transfer rates. $c = 500$ g/m$^3$ and $d_p = 1$ μm. LSF: lead shock front; RF: reaction front; MS: Mach stem; IW: incident wave.

**4.5 Interphase coupling**

The influences of coal particles on gaseous methane detonation are realized through mass, momentum, and energy exchanges and the corresponding source/sink terms are given in Eqs. (22)-(25). Figure 12 shows the instantaneous distributions of mass ($S_m$), energy ($S_e$) and momentum ($S_{mom}$) transfer rates when the DW propagates in the particle suspension. Here $c = 500$ g/m$^3$ and $d_p = 1$ μm. A positive mass (energy and momentum) transfer rate indicates that the corresponding transfer is from solid (gas) phase to gas (solid) phase. In the current modelling, interphase heat transfer includes the combined contributions from the convective heat transfer (gas → particle for particle heating) and char combustion heat release (particle → gas), as shown in Eq. (24). One can see in Fig. 12 that in the induction zone between the shock front LSF and reaction front RF, $S_e < 0$. This means that strong energy absorption occurs due to convective heat transfer for particle heating, which would weaken the detonation wave. Nonetheless, gradually increased heat is released from char burning in the induction zone and surface reaction rate is high around the RF. They are featured by high mass transfer rate $S_m$ in Fig. 12(a). Meanwhile, $S_{mom} < 0$ in the induction zone can be observed from Fig. 12(c), which indicates that there is a strong momentum transfer from the gas to accelerate the dispersed coal particles. This, to some degree, would also weakens the lead shock wave [33,45]. Nonetheless, the momentum transfer rate spatially decays quickly since the kinematic equilibrium is reached between two phases.

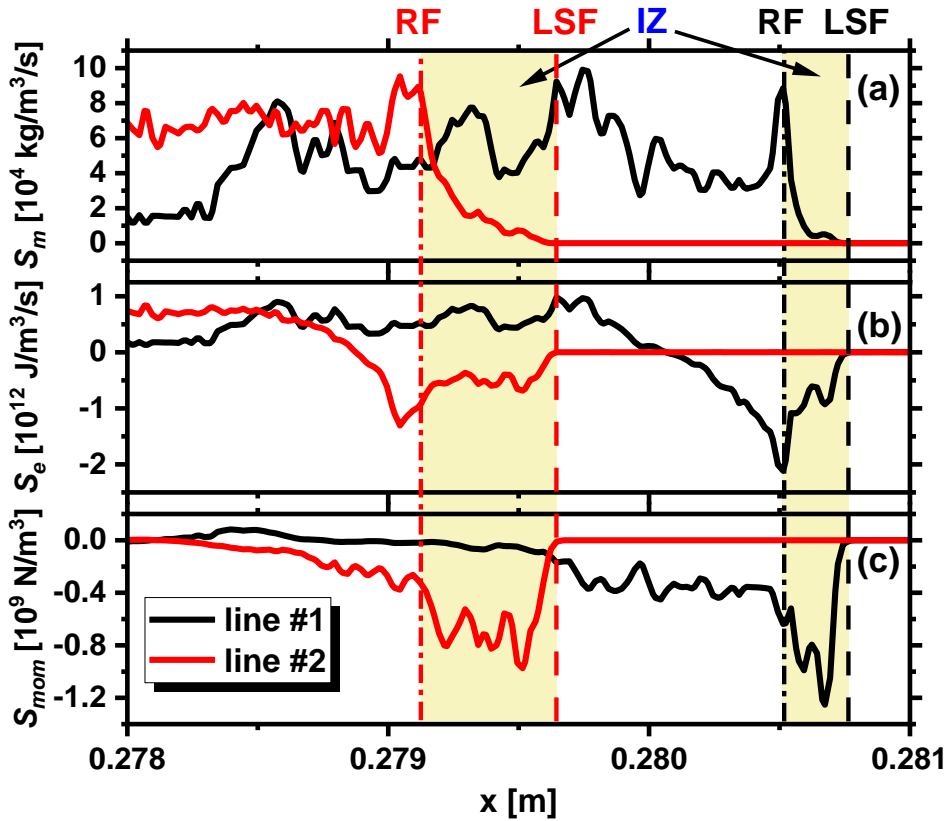

Fig. 13. Spatial profiles of the transfer rates of (a) mass, (b) energy, and (c) momentum across the Mach stem and incident wave. $c = 500$ g/m$^3$ and $d_p = 1$ μm. Lines #1 and #2 are marked in Fig. 11(c). IZ: induction zone.

To further quantify the two-phase coupling, the profiles of mass, energy and momentum transfer rates at lines #1 and #2 (marked in Fig. 12c) are shown in Fig.13. Lines #1 and #2 lie at a Mach stem and incident wave, respectively. The induction zone lengths along lines #1 and #2 are 510 and 240 µm, respectively, much lower than the HRL (2,200 µm) from the ZND structure of the stoichiometric $CH_4/O_2/N_2$ mixture. This may be mainly because the existence of char particles shortens the ignition time of gas phase, as will be discussed Section 4.7. According to Figs. 13(a) and 13(b), $S_m$ and $S_e$ increase gradually to the maximum near the RF, within the respective induction zones of lines #1 and #2. This indicates that the char particles absorb heat in the induction zone to heat themselves. The $S_m$ is highest around the RF because of high gas temperature there. To re-iterate, since $S_e$ is net rate of different energy transfer mechanisms (e.g., char combustion heat release), the heat actually absorbed by the particles in the IZ would be higher than what is shown in Fig. 13(b). Moreover, in Fig. 13(c), a strong momentum transfer mainly occurs immediately the lead SF, but decays very quickly beyond the induction zone.

To examine the two-phase coupling under different particle concentrations, time history of the averaged transfer rates of mass, energy and momentum is presented in Fig. 14. The results are density-weighted averaged interphase transfer rates ($S_m$, $S_{mom}$ and $S_{energy}$ in Eqs. 18-20) in the two-phase section (0.2-0.3 m, see Fig. 1). Five particle concentrations are considered in Fig. 14, which correspond to the same cases in Figs. 4 and 5. It can be seen from Fig. 14(a) that after the DW enters the coal particle suspension, the mass transfer rate $S_m$ increases gradually with time. In general, higher particle concentration results in larger $S_m$. However, for $c$ = 500 and 1,000 g/m³, as the DW extinction happen, the $S_m$ is lower than that of 250g/m³ for a period of time (> 9 µs). For the energy transfer rate, it is negative for the first several microseconds, and then changes to positive values. Besides, the greater the concentration, the greater the magnitude of the momentum transfer rate. As shown in Fig. 14(c), $S_{mom}$ rapidly increases to its maximum value when the DW enters the two-phase section, and then almost levels off. Meanwhile, the magnitude of $S_{mom}$ almost monotonically increases with the concentration.

Some observations related to detonation dynamics from Fig. 14 are worthy of further discussion. Firstly, for the 500 and 1000 g/m³ cases, when the DW just arrives at the suspension, the magnitudes of $S_m$ and $S_{energy}$ are high, and therefore the energy/momentum absorption effects would be high. This directly leads to the RF/LSF decoupling of near the leading edge of the suspension, as shown in Fig. 11. Secondly, the three transfer rates have peak values at the re-initiation instant. For instance, when $c = 1000$ g/m$^3$, they peak at around 30 μs. The multiple localized explosion pockets (see Figs. 9 and 10) significantly intensify the non-equilibria of the two-phase flows, which however decays to lower values in 5-10 μs in these cases.

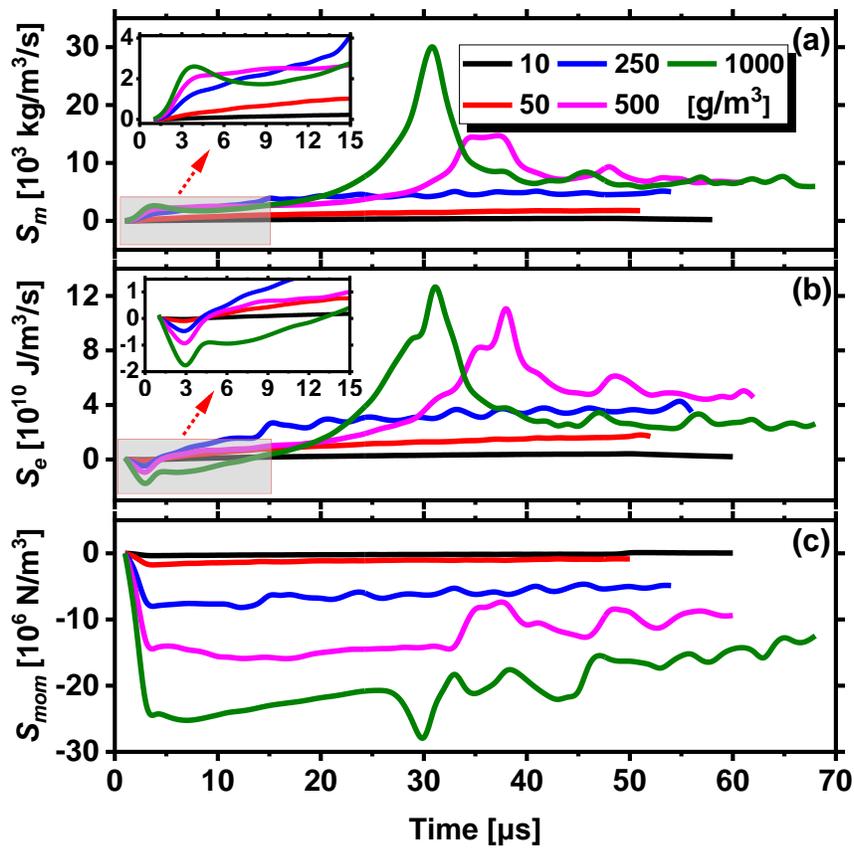

Fig. 14 Time history of the averaged transfer rates of (a) mass, (b) energy, and (c) momentum with different particle concentrations. $d_p = 1$ μm.

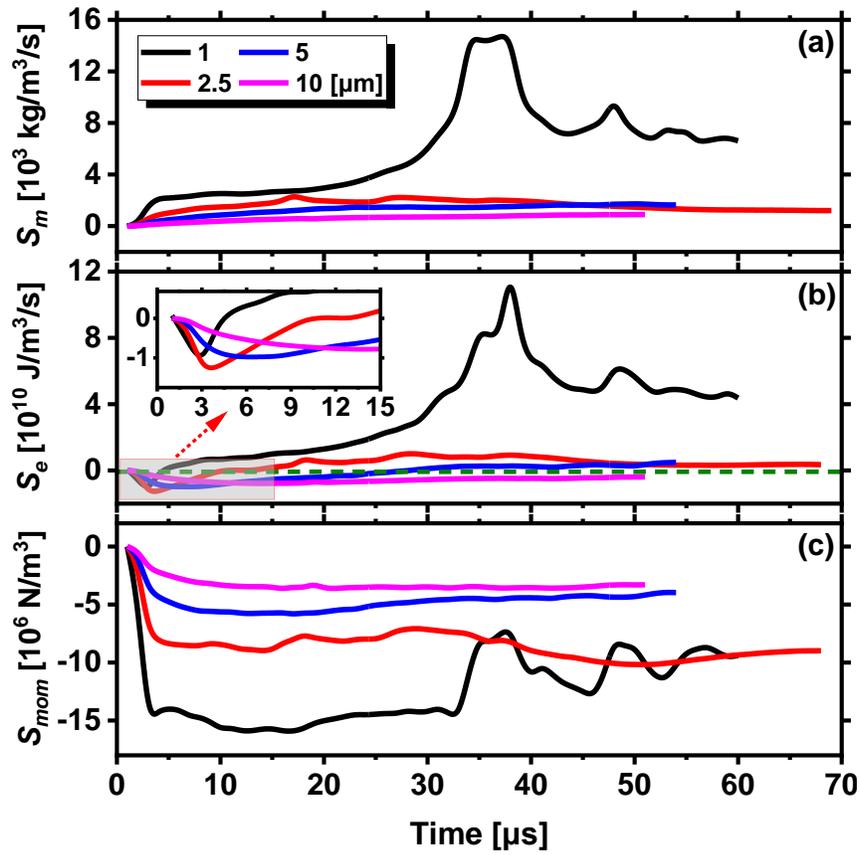

Fig. 15 Time history of the averaged transfer rates of (a) mass, (b) energy, and (c) momentum with different particle diameters. $c$= 500 g/m$^3$.

Likewise, the particle size effects on the two-phase coupling are shown in Fig. 15. We consider the particle diameters of 1-10 μm, corresponding to the same cases in Figs. 6 and 7. One can see from Fig. 15(a) that the smaller the particle size, the greater the transfer rates of mass. This suggests that small particles are more prone to ignite and burn [46]. However, after 45 μs, the $S_m$ of 2.5 μm particles is slightly lower than that of 5 μm ones, because the DW is decoupled in the former case. For $S_e$, the same phenomenon was observed after 52 μs. However, for $d_p$ = 5 μm, the energy transfer rate transits from negative value to positive one around 25 μs. This indicates that the larger the particle size, the less heat is released by the surface reaction of coal particles. Differently, for 10 μm particles, the energy transfer rate is always negative after the DW enters the two-phase section. This is because for coarser particles, the heat released by the surface reactions is much lower than the heat absorbed from the gas phase. This phenomenon will be further analyzed in section 4.7. It can be seen from the Fig. 15(c) that the smaller the particle size, the greater the momentum transfer rate. This is related to the faster response to the gas flows of the finer particles. For $d_p = 1$ μm, there is a significant fluctuation in the momentum transfer rate due to the re-initiation of the DW at about 35 μs.

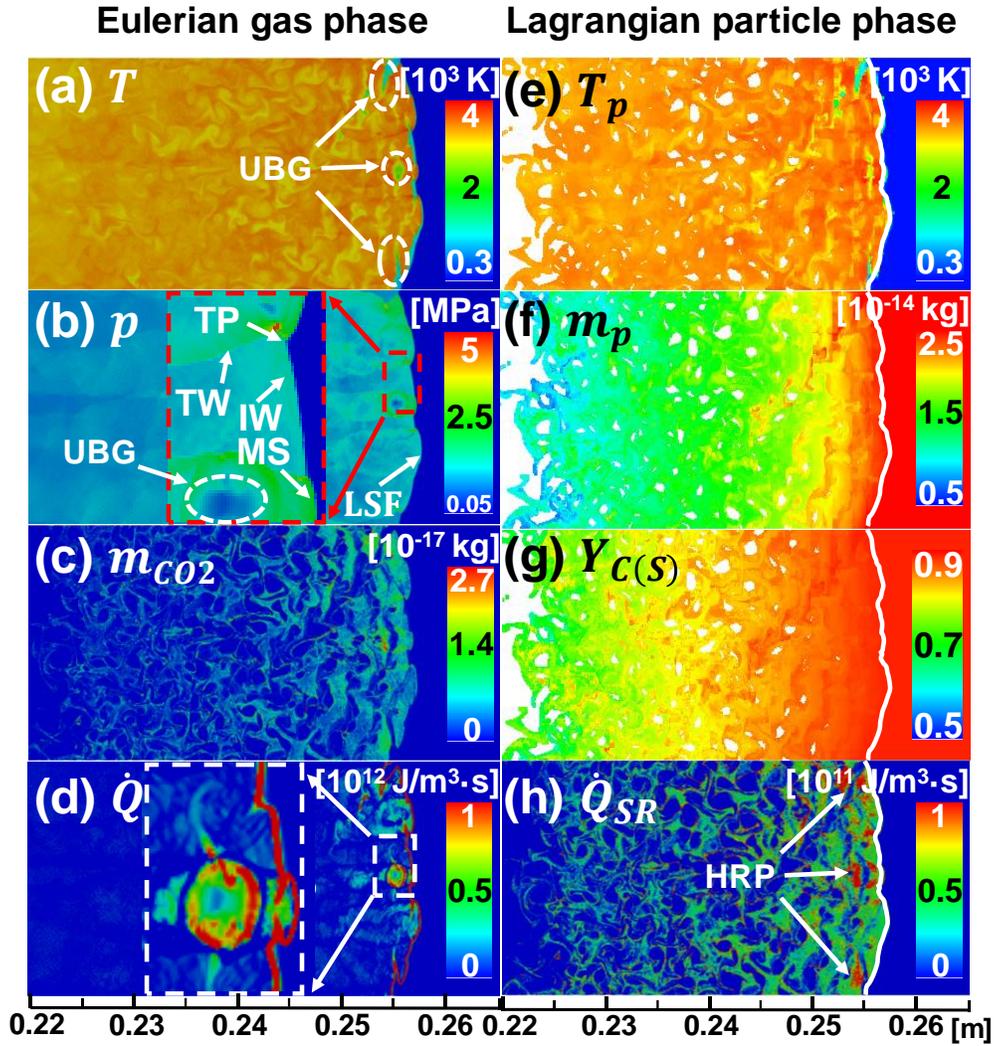

Fig. 16. Distributions of (a) gas temperature, (b) pressure, (c) $CO_2$ mass from surface reaction, (d) gas reaction heat release rate, (e) particle temperature, (f) particle mass, (g) carbon mass fraction in the particle, (h) surface reaction heat release rate. $c = 50$ g/m$^3$ and $d_p = 1$ μm. MS: Mach stem, TP: triple point, IW: incident wave, TW: transverse wave, UBG: unburned gas, HRP: heat release point. White line: shock front.

## 4.6 Hybrid detonation structure

Detailed structures of methane/coal particle hybrid detonation will be analyzed in this section. The gas and particle results with $c$ =50 g/m$^3$ and $d_p$=1 μm (same as Fig. 4c) are selected for analysis in Fig. 16. A weakly unstable detonation wave is observed, and the Mach stem, incident wave, transverse wave and triple point can be identified, as annotated in Fig. 16(b). The gas reaction heat release rate $\dot{Q}$ is high immediately behind the LSF, as shown in Fig. 16(d). Moreover, unburned gas (UBG) pockets exist in the detonation products (see Figs. 16a and 16b), and unburned mixtures are leaked behind the incident wave. One can see from Fig. 16(e) that the coal particles are heated to above 3,000 K immediately behind the LSF. This is reasonable because 1 μm particles have fast heating rate. Accordingly, the char starts to burn, and considerable CO$_2$ is produced from the surface reactions behind the Mach stems and incident waves (see Fig. 16c). This leads to quick reduction of coal particle mass $m_p$, evidenced in Fig. 16(f). Within 0.01 m behind the LSF, the mass of most particles is reduced to around 50% of the original value. In Fig. 16(g), the carbon mass fraction in the particles, $Y_{C(s)}$, is reduced to approximately 70% (but still not burned out yet) at 0.02 m behind the LSF.

Striped distributions of heat release from char combustion $\dot{Q}_{SR}$ can be found in Fig. 16(h). Several locations with high $\dot{Q}_{SR}$ can be seen (marked as HRP), which are caused by enhanced char combustion facilitated by the availability of the oxidant species in the unburned gas pockets. The localized strong surface reaction heat generation further promotes the homogeneous gas reactions, thereby higher $\dot{Q}$ near there (see Fig. 16d inset), which further elevates the local pressure. These pockets with char burning would be conducive for pressure wave formation, thereby affecting the lead shock.

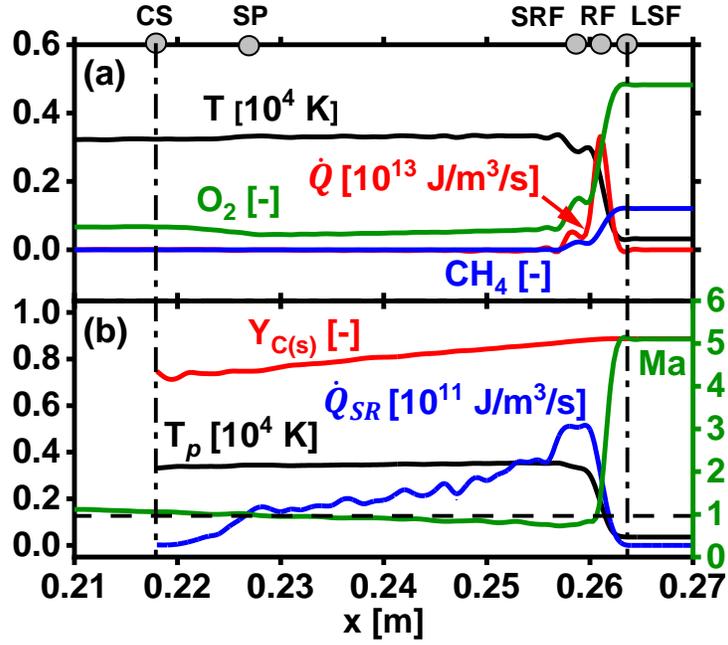

Fig. 17. Distributions of averaged (a) gas phase variables and (b) particle phase variables corresponding to the results in Fig. 16. $c = 50$ g/m$^3$ and $d_p = 1$ μm. LSF: lead shock front; RF: gas reaction front; SRF: surface reaction front; SP: shock-frame sonic point; CS: two-phase contact surface.

The structure of the hybrid detonation can be further quantified through averaging the key gas (density-weighted averaging) and particle (simple averaging) variables along the domain width (i.e., $y$-direction) and the results are presented in Fig. 17. At this instant, the $x$-direction length of the particle-laden area behind the lead shock is about 0.046 m, and the end of this area is a multiphase contact surface (CS). As observed from Fig. 17(a), the mean gas reaction HRR $\dot{Q}$ increases quickly after the shock and peaks around 0.26 m (termed as reaction front, RF). As such, the average induction distance between LSF and RF is about 3 mm. Accordingly, the mass fractions of CH$_4$ and O$_2$ quickly drop to around 0 and 0.067 respectively behind the reaction zone. The residual O$_2$ provides favorable environment for char combustion to proceed. One can see from Fig. 17(a) that the gas temperature $T$ rises rapidly to over 3,000 K due to detonative combustion, and the particle temperature $T_p$ (see Fig. 17b) basically follows the gas one due to the fast heating process. The maximum $\dot{Q}_{SR}$ (the corresponding location termed as SRF) lies slightly behind the RF. Nonetheless, continuous combustion of the coal particles leads to distributed char combustion HRR in the detonation products. From the distributions of the shock-frame Mach number $Ma$, the subsonic (actually very close to the sonic condition, like a C-J detonation) region spans from $x = 0.23$ to 0.26 m. The location of $Ma = 1$ corresponds to the sonic point (SP), i.e., $x = 0.23$ m in this structure. Therefore, char combustion largely proceeds in the subsonic region, which enables the influence of forward-running pressure waves from char combustion heat release on the lead shocks. The skeletal structure of the hybrid detonation, featured by foregoing key locations, is marked along the top $x$-axis in Fig. 17.

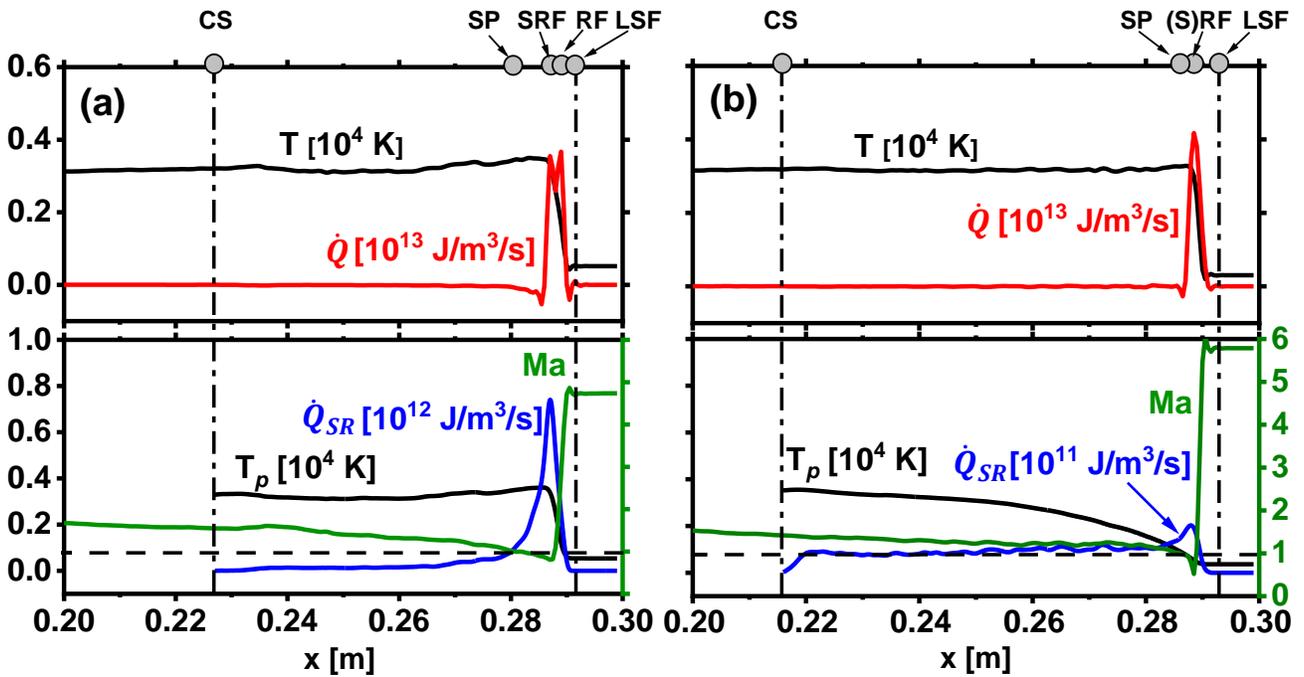

Fig. 18. Distributions of averaged variable corresponding to the results in Fig. 6: (a) 1 μm and (b) 10 μm. $c = 500$ g/m³.

How the hybrid detonation structure evolves with particle diameter or concentration merits further discussion. The distributions of averaged variable with different coal char particle diameters (1 and 10 μm) are shown in Fig. 18. The particle concentration is fixed to be $c = 500$ g/m³. For Fig. 18(a), the concentration $c$ is ten times higher than that in Fig. 17. It can be seen from Fig. 18(a) that the $x$-direction length of the particle-laden area behind the lead shock is about 0.066 m. The average induction distance between LSF and RF is about 1 mm. Compared with Fig. 17, when the concentration increases, the particle-laden area also increases, and the induction length decreases. This is because the larger the particle concentration is, the longer it takes for the particles to complete the surface reaction, and the stronger the weakening effect on the leading shock wave is due to the energy/momentum absorption of the particles. Moreover, the gas temperature $T$ and particle temperature $T_p$ rise rapidly to about 3200 K and 3400 K at a distance of 2 mm behind the LSF. This is close to that in Fig. 17 since the same particle size is considered. Furthermore, the maximum $\dot{Q}_{SR}$ in the 500 g/m³ case is one order of magnitude higher than that of the 50 g/m³ case, due to higher particle concentration. Accordingly, the peak gas reaction HRR, $\dot{Q}$ , are 0.33 and 0.38×10¹³ J/m³/s for 50 and 500 g/m³, respectively. The enhanced $\dot{Q}$ peak value may be associated with the stronger char combustion heat release in the latter case.

It can be seen from Fig. 18(a) the Mach number distribution that the range of subsonic propagation is about 7 mm, and the Mach number $Ma$ at the LSF is 4.8. The counterpart results of the $c = 50$ g/m³ case in Fig. 17 are 0.034 m and 5.1, respectively. Apparently, those from the 500 g/m³ case are much lower. This is reasonable, because of the weaker lead shock attenuated by higher-concentration particles.

Comparing Figs. 18(a) and 18(b) can indicate the particle size influences. When $d_p$= 10 μm in Fig. 18(b), the particle temperature $T_p$ increases monotonically towards the two-phase contact surface, due to smaller heating rate of larger particles. The peak gas reaction HRR $\dot{Q}$ is at the same position as that of char combustion $\dot{Q}_{SR}$. However, $\dot{Q}_{SR}$ is one order of magnitude smaller than that of the 1 μm particles. Meanwhile, the shock Mach number is about 5.8, and the subsonic zone length is only 4 mm. This implies that the larger the particle size is, the less the energy/momentum absorbed by the particles is, and the weakening effect on the leading shock is smaller.

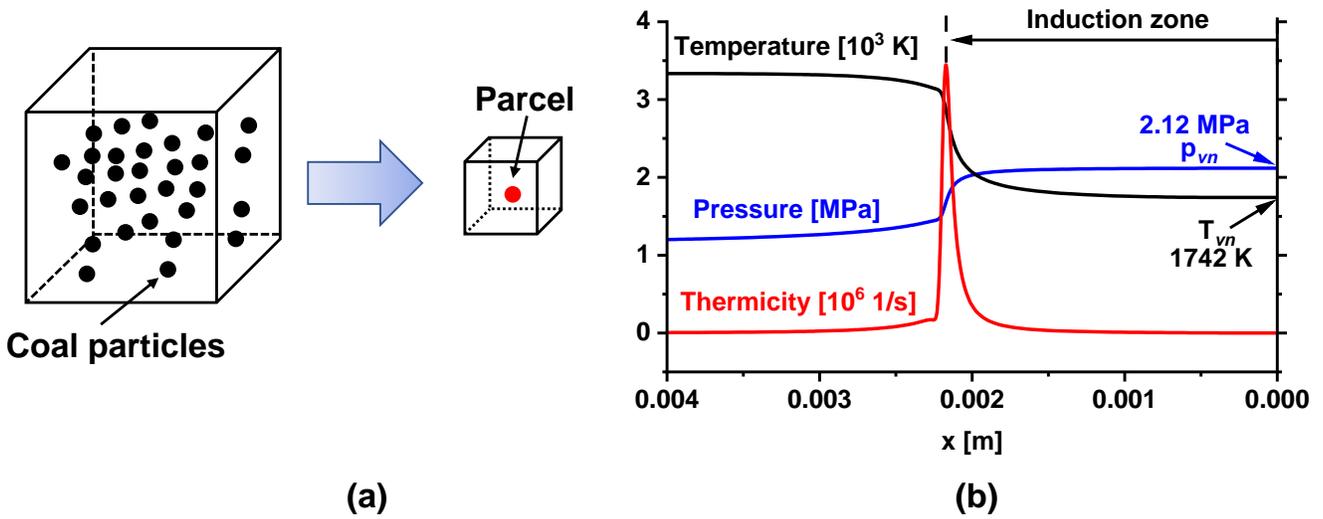

Fig. 19. (a) Domain of the constant-volume ignition with coal char particles. (b) The ZND profiles of pressure, temperature, and thermicity in stoichiometric $CH_4/O_2/N_2$ mixture.

**4.7 Discussion**

As seen from Figs. 17 and 18, the coal particles can burn in the induction zone and the reaction rate increases quickly near the RF due to the elevated gas temperature. It is necessary to further make fundamental discussion on how the coal particle burning affects the methane chemistry in the induction zone. Essentially, coal particles interact with the gas phase through heat and mass transfer. To investigate the influences of these interactions on gas phase chemistry, constant-volume ignition will be conducted in this section. We consider a cubic domain of $1 \times 1 \times 1$ mm$^3$, which is schematically shown in Fig. 19(a). The initial composition of the background gas is the same as that in 2D simulations, i.e., stoichiometric $CH_4/O_2/N_2$ mixture. The initial pressure and temperature are respectively taken as the von Neumann states of the ZND structure of the background gas, i.e., $T_{VN}$ =1,742 K and $p_{VN}$ = 2.12 MPa, shown in Fig. 19(b). Moreover, mono-sized coal particles of a given concentration are uniformly distributed in the domain, see Fig. 19(a). Consistent with the 2D simulations, the studied initial particle diameter and concentration are $d_p^0 = 0.25 - 30$ μm and $c = 10 - 1000$ g/m$^3$, respectively. The initial particle temperature is 300 K. The particle heat capacity and initial material density are the same as those in the 2D simulations.

One cell is used for the domain, and wall conditions are enforced for the six boundaries. The autoignition process is solved through Eqs. (1), (3), and (4) without the physical transport terms and radiation term. The computational parcel concept [41] is adopted, which groups all the particles with identical properties (e.g., temperature and mass). In our studies, one parcel is used to represent all the coal particles in the domain, which is placed at the cell center, see Fig. 19(a). This essentially corresponds to a 0D ignition calculation with mass and heat exchanges between the gas and coal particles. These calculations can provide insightful results about the dispersed phase on gas chemistry, e.g., in Ref. [47].

Ignition delay time (IGT) of the gas mixture as a function of particle diameter and concentration are calculated, and the results are shown in Figs. 20 and 21. Here the IGT is defined as the duration from the beginning to the instant with maximum HRR. The IGT of the particle-free stoichiometric $CH_4/O_2/N_2$ mixture, 6.45 µs, is also added for comparison (termed as "gas IGT" hereafter).

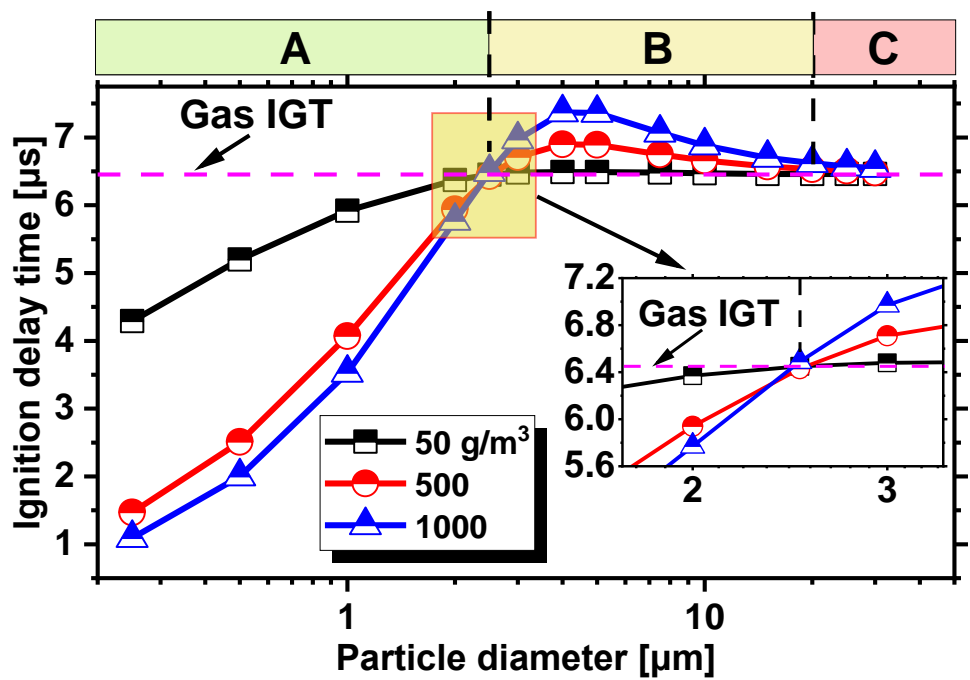

Fig. 20. Ignition delay time as a function of particle diameter with different particle concentrations.

As demonstrated in Fig. 20, when the particle diameter $d_p^0$ is around 2.5 μm, the IGT of the two-phase mixture is close to that of the particle-free case. Meanwhile, when $d_p^0$ is beyond a certain value, i.e., 20 μm, the IGT also approaches the gas IGT value. These tendencies exist for all three concentrations. Therefore, with the foregoing two critical diameters, the dependence of IGT on particle diameter can be divided into Regimes A, B and C. Specifically, in Regime A ($d_p^0 < 2.5$ μm), the ignition is earlier than that in the gas-only mixture. This is because the heat released by the small particle surface reactions facilitates the gas phase reaction. Moreover, the smaller the particle size, the shorter the IGT, as demonstrated in Fig. 20. In Regime B ($2.5 < d_p^0 < 20$ μm), the IGT is higher than the gas IGT. With increased particle size, the IGT first increases and then decreases. This non-monotonicity may be induced by the competition between particle heat absorption from the gas phase and heat release due to char combustion. In Regime C, when the particle size further increases beyond 20 μm, the IGT is not sensitive to the particle diameter variations, which is close to the gas IGT.

Interestingly, in each regime, particle concentration exhibits different effects on IGT. Figure 21 further demonstrates the relations between the IGT and particle concentration. One can see that, for a fixed particle size (e.g., 1 or 10 μm), the IGT always monotonically changes with the particle concentration. This means that the larger the particle concentration, the greater the influence of coal particle on the IGT. Nonetheless, the exceptions are the results for the critical diameter of 2.5 μm and the case in Regime C, which shows weak dependence on particle concentration. This may be due to the reason that the heat absorbed by the particles is close to the one from char burning. In Regime A, for small particles, e.g., 1 and 2 μm, the IGT gradually decreases when $c$ is increased. This is because the higher concentration, the more heat is absorbed to heat the particles. This can also further justify the unsteady phenomena in detonations with micro-sized particles in Figs. 4(e) and 4(f). For 5 and 10 μm coal particles, IGT increases with particle concentration.

Based on the results in Figs. 20 and 21, one can see that the effects of the coal particles are multi-fold. Micron-sized or submicron coal particles can kinetically facilitate the gas chemistry. Therefore, they are more hazardous to explosion, e.g., in coal mines. However, they would be preferably used for coal particle co-burning in propulsion system. Particles of these sizes are expected to ignite and burn within the hydrodynamic length of the detonation [46]. For coarse particles, their effects on the gas chemical kinetics are relatively weak.

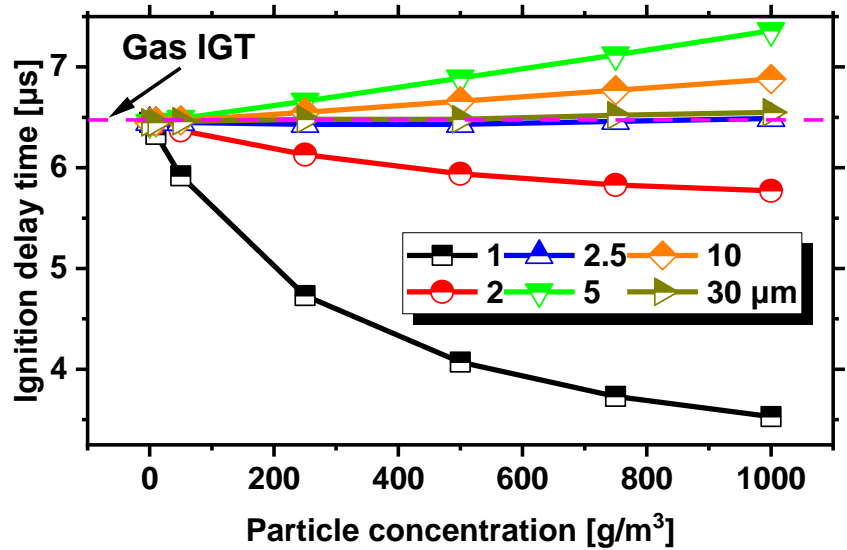

Fig. 21. Ignition delay time as a function of particle concentration with different initial particle diameters.

The different effects of dispersed coal particles on IGT can be justified by the interactions between the gas phase reaction and char combustion. In Fig. 22, we select one typical case from each regime for analysis. The results (dotted line) from the particle-free case are added for comparison. For $d_p^0 = 1$ μm, in Figs. 22(a) and 22(b), the particles are heated by the hot gas and the temperature increases from the initial one to 2,000 K around 1 μs. From 1 to 4 μs, the coal particle temperature gradually increases towards over 3,000 K, accompanied by the char combustion with finite heat release. During this period, the particle temperature is higher than the gas temperature and therefore particle would in turn heat the gas phase, which is in the chemical runaway stage. At 4 µs, thermal runaway of the gaseous mixture happens, with strong heat release and temperature rise. Therefore, the char combustion and gas chemistry have pronounced interactions, and the gas HRR is about one order of magnitude higher than that of char combustion (black line in the lower panel).

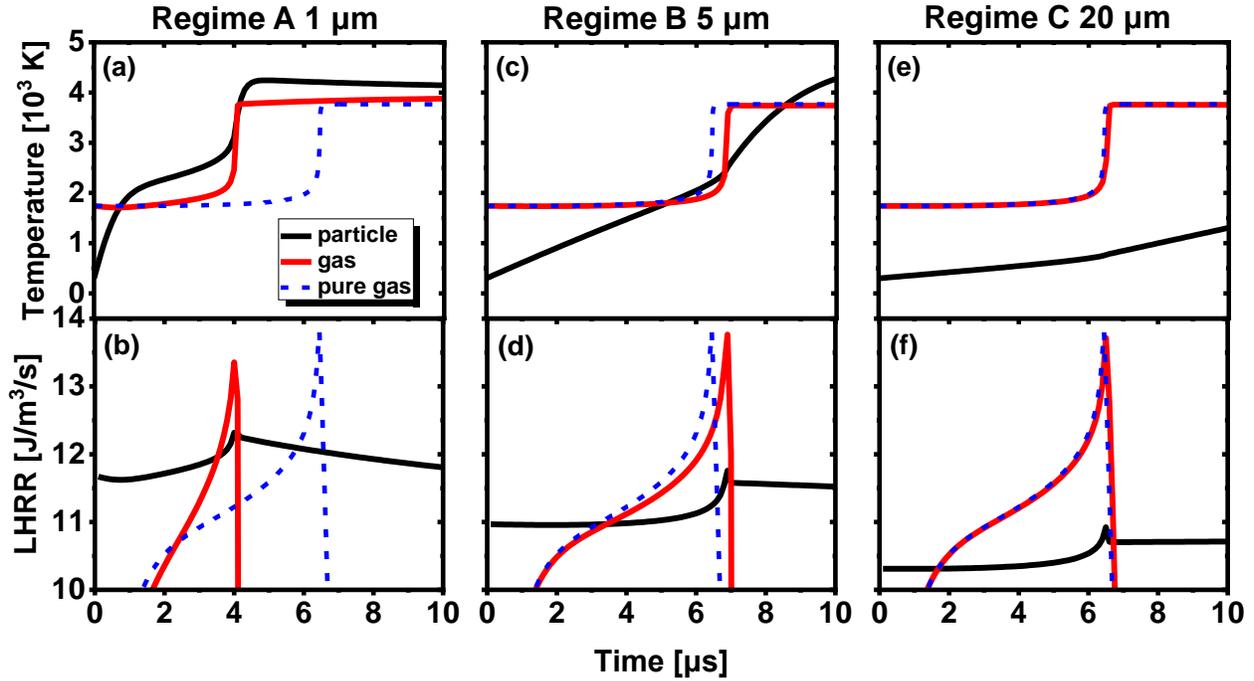

Fig. 22. Time history of gas and particle temperature, gas and particle reaction heat release rate with different particle diameters in the different regimes. $c$= 500 g/m$^3$, $LHRR \equiv sign(HRR) \cdot \log_{10}(1 + |HRR|)$.

For the case in Regime B, due to larger particle size, heating rate is much lower than that in Fig. 22(a). As such, the particle temperature is lower than the gas temperature for the majority of the chemical runaway stage. Due to this temperature difference, heat absorption from the background gas happens and the gas chemistry would be delayed due to the existence of the coal char particles. The char combustion heat release is also much lower than the 1 µm case. Therefore, for Regime B, the auto-ignition time of the gas-phase reaction is prolonged. When the particle size further increases, e.g., $d_p^0 = 20$ µm, it is also seen from Figs. 22(e) and 22(f) that during the induction period of gas chemical reactions, the particle temperature rises even more slowly. The gas temperature is much higher than the particle temperature. Therefore, for Regime C, the ignition history is almost not affected by the particles.

To further analyze the surface reaction effects under various particle conditions, numerical experiments are performed by turning off the surface reaction model in the simulations, and the comparisons of their IGTs are presented in Fig. 23 for two concentrations of 50 and 1000 g/m$^3$. One can see that, for either concentration, the surface reaction only shows a considerable influence when the particle size is smaller than a certain value, i.e., around 5 µm for the two concentrations. This further identifies particle size range within which the surface reaction would exhibit a substantial influence on gas phase chemistry, thereby justifying the rich detonation behaviors (such as localized explosion and detonation re-initiation) with small particles seen in the 2D simulations.

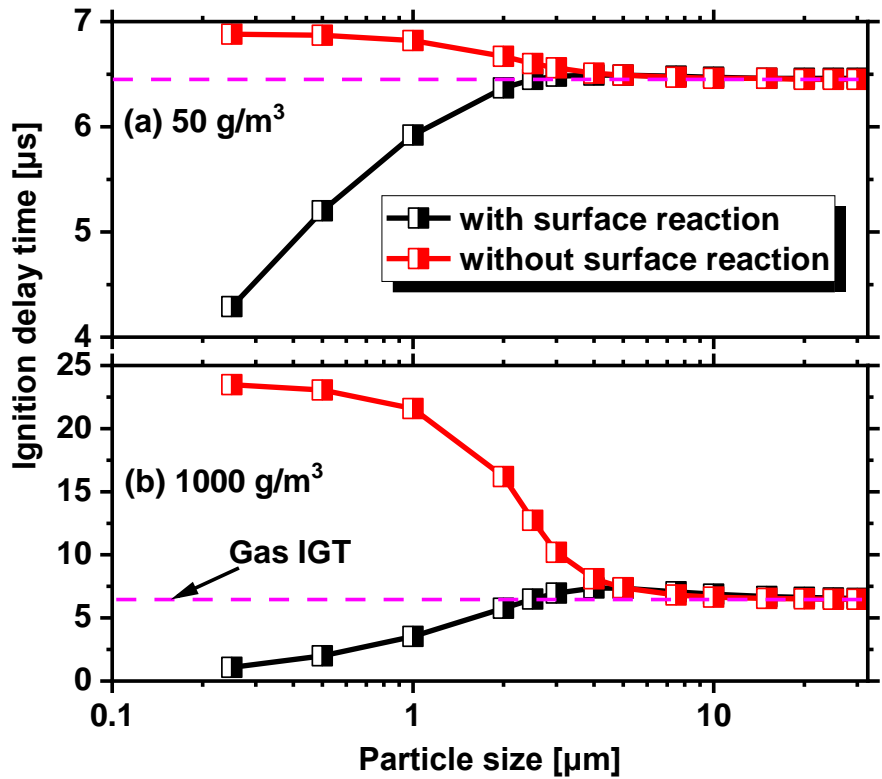

Fig. 23. Effects of surface reaction on ignition delay time of gas phase mixture with different particle sizes.

## 5. Conclusions

Methane detonation dynamics in dilute coal char particle suspensions are computationally studied with a Eulerian-Lagrangian appraoch. Two-dimensional configuration is consdiered and a reduced chemical mechanism is employed. Parametric studies are performed with different particle sizes and concentrations. Key conclusions are summarized as below:

The methane detonation wave propagation is considerably affected by coal particle concentration and size. Detonation extinction occurs near the leading edge of the suspension when the particle size is small and concentration is high. The averaged lead shock speed decreases with increased particle concentration and decreased particle size.

Moreover, for 1 μm particle, if the particle concentration is beyond a threshold value (465 g/m$^3$), detonation re-initiation occurs. This is caused by the shock focusing along the reaction front in a decoupled detonation and these shocks are generated from surface reactions behind the lead shock. A regime map of detonation propagation and extinction is predicted. It is found that the re-initiation location decreases with particle concentration, but approaches a constant (~0.225 m) when the concentration exceeds 1000 g/m$^3$.

In addition, the interphase coupling between the detonation wave and coal particle are discussed. The mass and energy transfer rate increase rapidly to the maximum near the reaction front in the induction zone. Meanwhile, the smaller the particle size and the larger the particle concentration, the greater the transfer rates of mass, energy, and momentum.

Detailed structures of methane/coal particle hybrid detonation are also studied. The results reveal that the several locations with high heat release are caused by enhanced char combustion facilitated by the availability of the oxidant species in the unburned gas pockets. These pockets with char burning would be conducive for pressure wave formation, thereby affecting the lead shock. The one-dimensionalized structures are also analyzed with various key locations, including sonic point, two-phase contact surface, and reaction fronts from gas chemistry and particle surface reactions. The particle properties have significant effects on the hybrid detonation structures.

Finally, the influence of coal particle surface reaction on methane chemistry is studied based on constant-volume ignition calculations. It is found that the surface reaction has significant effects on IGT when particle size is less than 2.5 μm. Moreover, the IGT changes non-monotonically with particle size. The dependence of IGT on particle diameter can be divided into Regimes A, B and C. Specifically, in the Regime A ($d_p^0 < 2.5$ μm), the ignition is earlier than that in the gas IGT. In Regime B ($2.5 < d_p^0 < 20$ μm), the IGT is higher than that of the gas IGT. In Regime C, when the particle size further increases beyond 20 μm, the IGT is almost unaffected when the particle size varies, close to the IGT of gaseous mixture. Moreover, the IGT monotonically changes with the particle concentration.

**Acknowledgements**

The numerical simulations were performed with the computational resources from National Supercomputing Center, Singapore (NSCC, https://www.nscc.sg/) and the Fugaku Supercomputer in Japan (hp210196). JS is funded by the China Scholarship Council (202006420042). WR is supported by "the Fundamental Research Funds for the Central Universities" (2019XKQYMS75). HZ is supported by MOE Tier 1 Grant (A-0005238-00-00). The ZND profiles are computed with The Shock and Detonation Toolbox developed by Professor Joe Shepherd at Caltech.